\RequirePackage{ifpdf}
\ifpdf % We are running pdfTeX in pdf mode
\documentclass[pdftex]{sigma}
\else
\documentclass{sigma}
\fi

\numberwithin{equation}{section}

\def\p{\partial}

\begin{document}

\allowdisplaybreaks

\renewcommand{\thefootnote}{$\star$}

\renewcommand{\PaperNumber}{083}

\FirstPageHeading

\ShortArticleName{The Noncommutative Doplicher--Fredenhagen--Roberts--Amorim Space}

\ArticleName{The Noncommutative\\ Doplicher--Fredenhagen--Roberts--Amorim Space\footnote{This paper is a
contribution to the Special Issue ``Noncommutative Spaces and Fields''. The
full collection is available at
\href{http://www.emis.de/journals/SIGMA/noncommutative.html}{http://www.emis.de/journals/SIGMA/noncommutative.html}}}

\Author{Everton M.C. ABREU~$^{\dag\ddag}$, Albert C.R. MENDES~$^\S$, Wilson OLIVEIRA~$^\S$\\ and Adriano O. ZANGIROLAMI~$^\S$}

\AuthorNameForHeading{E.M.C.~Abreu, A.C.R.~Mendes, W.~Oliveira and A.O.~Zangirolami}

\Address{$^\dag$~Grupo de F\'{\i}sica Te\'orica e Matem\'atica F\'{\i}sica, Departamento de F\'{\i}sica, \\
\hphantom{$^\dag$}~Universidade Federal Rural do Rio de Janeiro,
BR 465-07, 23890-971, Serop\'edica, RJ, Brazil}
\EmailD{\href{mailto:evertonabreu@ufrrj.br}{evertonabreu@ufrrj.br}}

\Address{$^\ddag$~Centro Brasileiro de Pesquisas F\'{\i}sicas (CBPF), Rua Xavier Sigaud 150,\\
\hphantom{$^\ddag$}~Urca, 22290-180, RJ, Brazil}

\Address{$^\S$~Departamento de F\'{\i}sica, ICE, Universidade Federal de Juiz de Fora, \\
\hphantom{$^\S$}~36036-330, Juiz de Fora, MG, Brazil}
\Email{\href{mailto:albert@fisica.ufjf.br}{albert@fisica.ufjf.br},
\href{mailto:wilson@fisica.ufjf.br}{wilson@fisica.ufjf.br}, \href{mailto:adrianozangirolami@fisica.ufjf.br}{adrianozangirolami@fisica.ufjf.br}}

\ArticleDates{Received March 28, 2010, in f\/inal form October 02, 2010;  Published online October 10, 2010}

\Abstract{This work is an ef\/fort in order to compose a pedestrian review of the recently elaborated Doplicher, Fredenhagen, Roberts and Amorim (DFRA) noncommutative (NC) space which is a minimal extension of the DFR space.
In this DRFA space, the object of noncommutativity ($\theta^{\mu\nu}$) is a variable of the NC system and has a  canonical conjugate momentum.  Namely, for instance, in NC quantum mechanics we will show that $\theta^{ij}$ $(i,j=1,2,3)$ is an operator in Hilbert space and we will explore the consequences of this so-called ``ope\-ra\-tionalization''.  The DFRA formalism is constructed in an extended space-time with independent degrees of freedom associated with the object of noncommutativity~$\theta^{\mu\nu}$.
We will study the symmetry properties of an extended $x+\theta$ space-time, given by the group~${\cal P}'$, which has the Poincar\'{e} group ${\cal P}$ as a subgroup.  The Noether formalism adapted to such
extended $x+\theta$ $(D=4+6)$ space-time is depicted.
A consistent  algebra involving the enlarged set of canonical operators is described, which permits one to construct theories that are dynamically invariant under the action of the rotation group.
In this framework it is also possible to give dynamics to the NC operator sector, resulting in new  features.
A~consistent classical mechanics formulation  is analyzed in such a way that, under quantization, it
furnishes a~NC quantum theory with interesting results.
The Dirac formalism for constrained Hamiltonian systems is considered
and  the object of noncommutativity  $\theta^{ij}$ plays a fundamental role as an~independent quantity.
Next, we explain the dynamical spacetime symmetries   in NC rela\-tivistic theories by using the DFRA algebra.
It is also explained about the generalized Dirac equation issue, that the fermionic f\/ield depends not only  on the ordinary coordinates but  on~$\theta^{\mu\nu}$ as well.  The dynamical symmetry content of such  fermionic theory is discussed, and we show that its action is invariant under~${\cal P}'$.
In the last part of this work we analyze the complex scalar f\/ields using this new framework.
As said above, in a~f\/irst quantized formalism, $\theta^{\mu\nu}$
and its canonical momentum $\pi_{\mu\nu}$ are seen as operators  living in some Hilbert space.
In a~second quantized formalism perspective, we show an explicit form for the extended Poincar\'e
generators and the same algebra is generated via generalized Heisenberg relations.  We also consider a source term and construct the general solution for the complex scalar f\/ields using the
Green function technique.}


\Classification{70S05; 70S10; 81Q65; 81T75}

\tableofcontents

\renewcommand{\thefootnote}{\arabic{footnote}}
\setcounter{footnote}{0}

\section{Introduction}

Theoretical physics is living nowadays a moment of great excitement and at the same time great anxiety before the possibility by the Large Hadron Collider (LHC) to reveal the mysteries that make the day-by-day of theoretical physicists.  One of these possibilities is that, at the collider energies, the extra and/or the compactif\/ied spatial dimensions become manifest.  These manifestations can lead, for example, to the fact that standard four-dimensions spacetimes may become NC, namely, that the position four-vector operator ${\mathbf x}^\mu$ obeys the following rule
\begin{gather} \label{a1}
[{\mathbf x}^\mu,{\mathbf x}^\nu]=i\theta^{\mu\nu},
\end{gather}
 where $\theta^{\mu\nu}$ is a real, antisymmetric and constant matrix.
The f\/ield theories def\/ined on a spacetime with (\ref{a1}) have its Lorentz invariance obviously broken.  All the underlying issues have been explored using the representation of the standard framework of the Poincar\'e algebra through the Weyl--Moyal correspondence.  To every f\/ield operator $\varphi({\mathbf{x}})$ it has been assigned a Weyl sym\-bol~$\varphi(x)$, def\/ined on the commutative counterpart of the NC spacetime.  Through this correspondence, the products of operators are replaced by Moyal $\star$-products of their Weyl symbols
\[
\varphi({\mathbf{x}})\psi({\mathbf{x}}) \longrightarrow \varphi(x) \star \psi(x),
\]
where we can def\/ine the Moyal product as
\begin{gather} \label{a11.22}
\varphi(x) \star \psi(x) = \exp\left[\frac{i}{2} \theta^{\mu\nu} \frac{\partial}{\partial x^{\mu}}\frac{\partial}{\partial y^{\nu}} \right] \varphi(x) \psi(y)\big|_{x=y}
\end{gather}
and where now the commutators of operators are replaced by Moyal brackets as,
\[
[x^\mu,x^\nu ]_{\star}\equiv x^\mu\star x^\nu  - x^\nu \star x^\mu  =  i \theta^{\mu\nu} .
\]
From (\ref{a11.22}) we can see clearly that at zeroth-order the NCQFT is Lorentz invariant.  Since $\theta^{\mu\nu}$ is valued at the Planck scale, we use only the f\/irst-order of the expansion in (\ref{a11.22}).

But it was Heisenberg who suggested, very early, that one could use a NC structure for spacetime coordinates at very small length scales to introduce an ef\/fective ultraviolet cutof\/f.
After that, Snyder tackled the idea launched by Heisenberg and published what is considered as the f\/irst paper on spacetime noncommutativity in 1947~\cite{Snyder}.
C.N.~Yang, immediately after Snyder's paper, showed that the problems of f\/ield theory supposed to be removed by noncommutativity were actually not solved~\cite{yang} and he tried to recover the translational invariance broken by Snyder's model.
The main motivation was to avoid singularities in quantum f\/ield theories.
However, in recent days, the issue has been motivated by string theory~\cite{Strings} as well as by other issues in physics \cite{other,Deriglazov,alexei,gangopadhyay}. For reviews in NC theory, the reader can f\/ind them in~\cite{QG,NCFT}
(see also~\cite{DFR}).

In his work Snyder introduced a f\/ive dimensional spacetime with ${\rm SO}(4,1)$ as a symmetry group, with generators ${\mathbf M}^{AB}$, satisfying the Lorentz algebra, where $A,B=0,1,2,3,4$ and using natural units, i.e., $\hbar=c=1$. Moreover, he introduced the relation between coordinates and  generators of the $SO(4,1)$ algebra
\[
{\mathbf x}^{\mu}=a {\mathbf M}^{4\mu}
\]
(where $\mu,\nu=0,1,2,3$ and the parameter $a$ has dimension of length), promoting in this way the spacetime coordinates to Hermitian operators. The mentioned relation introduces the commutator,
\begin{gather}
\label{2}
[{\mathbf x}^\mu,{\mathbf x}^\nu] = i a^2{\mathbf M}^{\mu\nu}
\end{gather}
and the identities,
\begin{gather*} %\label{2.1}
 [{\mathbf M}^{\mu\nu},{\mathbf x}^\lambda]= i\big({\mathbf x}^{\mu}\eta^{\nu\lambda}-{\mathbf x}^{\nu}\eta^{\mu\lambda}\big)
\end{gather*}
and
\begin{gather*} %\label{2.2}
 [{\mathbf M}^{\mu\nu},{\mathbf M}^{\alpha\beta}]= i\big({\mathbf M}^{\mu\beta}\eta^{\nu\alpha}-{\mathbf M}^{\mu\alpha}\eta^{\nu\beta}+{\mathbf M}^{\nu\alpha}\eta^{\mu\beta}-{\mathbf M}^{\nu\beta}\eta^{\mu\alpha}\big) ,
\end{gather*}
which agree with four dimensional Lorentz invariance.

Three decades ago Connes   et al.~(see~\cite{connes}  for a review of this formalism) brought the concepts of noncommutativity by generalizing the idea of a dif\/ferential structure to the NC formalism.  Def\/ining a generalized integration~\cite{connes2} this led to an operator algebraic description of NC spacetimes and hence, the Yang--Mills gauge theories can be def\/ined on a large class of NC spaces.  And gravity was introduced in~\cite{noncommut}.  But radiative corrections problems cause its abandon.

When open strings have their end points  on D-branes in the presence of a background constant B-f\/ield,  ef\/fective gauge theories on a NC spacetime arise~\cite{Hull,SW}. In these NC f\/ield theories (NCFT's)~\cite{NCFT},  relation~(\ref{2}) is  replaced by equation~(\ref{a1}).   A NC gauge theory originates from a~low energy limit for open string theory embedded in a constant antisymmetric background f\/ield.

The fundamental point about the standard NC space is that the object of noncommutativi\-ty~$\theta^{\mu\nu}$ is usually assumed to be a constant antisymmetric matrix in NCFT's. This  violates  Lorentz symmetry because it f\/ixes a direction in an inertial reference frame.   The violation of Lorentz invariance is problematic, among other facts, because it brings ef\/fects such as vacuum birefringence~\cite{Jaeckel}.  However, at the same time it permits to treat NCFT's as deformations of ordinary quantum f\/ield theories, replacing ordinary products with Moyal products, and ordinary gauge interactions by the corresponding NC ones. As it is well known, these theories carries serious problems as  nonunitarity, nonlocalizability, nonrenormalizability, $UV \times IR$  mixing etc.
On the other hand, the Lorentz invariance can be recovered by constructing the NC spacetime with
$\theta^{\mu\nu}$ being a tensor operator with the same hierarchical level as the ${\mathbf x}$'s. This was done in~\cite{Carlson} by using a convenient reduction of Snyder's algebra. As ${\mathbf x}^\mu$ and $\theta^{\mu\nu}$  belong in this case to the same af\/f\/ine algebra,  the f\/ields must be functions of the
eigenvalues of both ${\mathbf x}^\mu$ and $\theta^{\mu\nu}$. In~\cite{Banerjee2} Banerjee et al.\ obtained conditions for preserving Poincar\'e invariance in NC gauge theories and a whole investigation about various spacetime symmetries was performed.

The results appearing in~\cite{Carlson} are explored by some authors
\cite{Banerjee2,Morita,Haghighat,Carone,Ettefaghi,Saxell}. Some of them prefer  to start from the beginning  by adopting the Doplicher, Fredenhagen and Roberts (DFR) algebra~\cite{DFR}, which essentially assumes~(\ref{a1}) as well as the vanishing of the triple commutator among the coordinate operators.  The DFR algebra is based on  principles imported from general relativity (GR) and quantum mechanics (QM). In addition to~(\ref{a1}) it also assumes that
\begin{gather}
\label{6}
[{\mathbf x}^\mu,\theta^{\alpha\beta}] = 0 .
\end{gather}

With this formalism, DFR demonstrated that the combination of QM with the classical gravitation theory, the ordinary spacetime loses all operational meaning at short distances.

An important point in DFR algebra is that the Weyl representation of NC operators \mbox{obeying} (\ref{a1}) and~(\ref{6}) keeps the usual form of the Moyal product, and consequently the form of the usual NCFT's, although the f\/ields have to be considered as depending not only on~${\mathbf x}^\mu$ but also on~$\theta^{\alpha\beta}$.
The argument is that  very accurate measurements of spacetime localization could transfer to test particles energies suf\/f\/icient to create a gravitational f\/ield that in principle could trap photons. This possibility is related with spacetime uncertainty relations that can be derived from~(\ref{a1}) and~(\ref{6}) as well as from the quantum conditions
\begin{gather}
\theta_{\mu\nu}\theta^{\mu\nu} =0,\qquad
 \left({1\over4}\,{}^*\theta^{\mu\nu} \theta_{\mu\nu}\right)^2=\lambda_P^8,\label{04000}
\end{gather}
where $^*\theta_{\mu\nu}={1\over2}\epsilon_{\mu\nu\rho\sigma} \theta^{\rho\sigma}$ and $\lambda_P$ is the Planck length.

These operators are seen as acting on a Hilbert space ${\cal H}$ and this theory implies in extra compact dimensions~\cite{DFR}. The use of conditions (\ref{04000}) in~\cite{Carlson,Morita,Haghighat,Carone,Ettefaghi,Saxell} would bring trivial consequences, since  in those works the relevant results  strongly depend on the value of $\theta^2$, which is taken as a mean with some weigh function $W(\theta)$. They use in this process the celebrated Seiberg--Witten \cite{SW} transformations.
Of course those authors do not use (\ref{04000}), since their motivations are not related to quantum gravity  but basically with the construction of a NCFT which keeps Lorentz invariance. This is a fundamental  matter, since there is no experimental evidence to assume Lorentz symmetry  violation~\cite{Jaeckel}.  Although we will see that in this review we are not using twisted symmetries~\cite{twisted,Gracia,wess,wess2} there is some considerations  about the ideas and concepts on this twisted subject that we will make in the near future here.

A nice framework to study aspects on noncommutativity is given by the so called NC quantum mechanics (NCQM), due to its simpler approach. There are several interesting works in NCQM
\cite{Deriglazov,Durval,Chaichan,Chaichan2,Gamboa,Nair,Banerjee,Bellucci,Ho,Smailagic,Jonke,Kokado,Kijanka,Dadic,Bellucci1,Calmet,Scholtz,Rosenbaum}.
In most of these papers, the object of noncommutativity $\theta^{ij}$ (where $i,j=1,2,3$), which essentially is the result of the commutation of two  coordinate operators, is considered as a constant matrix, although this is not the general case~\cite{Snyder,Deriglazov,Carlson,Morita,DFR}. Considering $\theta^{ij}$ as a constant matrix spoils the Lorentz symmetry  or correspondingly the rotation symmetry for nonrelativistic theories.
In NCQM, although time is a commutative parameter, the space coordinates do not commute.  However, the objects of noncommutativity are not considered as Hilbert space operators. As a consequence the corresponding conjugate momenta is not introduced, which, as well known, it is important to implement rotation as a dynamical symmetry~\cite{Iorio}.  As a result, the theories are not invariant under rotations.

In his f\/irst paper \cite{Amorim1}, R.~Amorim promoted an extension of the DFR algebra to a non-relativistic QM in the trivial way, but keeping consistency.
The objects of noncommutativity were considered as  true operators and their conjugate momenta were introduced. This permits to display a complete and consistent algebra among the Hilbert space operators and to construct generalized angular momentum operators, obeying the $SO(D)$ algebra, and in a dynamical way, acting properly in all the sectors of the Hilbert space.  If this is not accomplished, some fundamental objects usually employed in the literature, as the shifted coordinate operator (see~(\ref{16000})), fail to properly transform under rotations. The symmetry is implemented not in a mere algebraic way, where the transformations are based on the indices structure of the variables, but it comes dynamically from the consistent action of an operator, as discussed in~\cite{Iorio}.  This new NC space has ten dimensions and now is known as the Doplicher--Fredenhagen--Roberts--Amorim (DFRA) space.  From now on we will review  the details of this new NC space and describe the recent applications published in the literature.

We will see in this review that a consistent classical mechanics formulation  can be shown in such a way that, under quantization, it
gives a  NC quantum theory with interesting new features. The Dirac formalism for constrained Hamiltonian systems is strongly used,
and  the object of noncommutativity~$\theta^{ij}$ plays a fundamental role as an independent quantity. The presented classical theory, as its quantum counterpart, is naturally invariant under the rotation group $SO(D)$.

The organization of this work is: in Section~\ref{section2} we describe the mathematical details of this new NC space.  After this we describe the NC Hilbert space and construct an harmonic oscillator model for this new space.  In Section~\ref{section3} we explore the DFRA classical mechanics and treat the question of Dirac's formalism.  The symmetries and the details of the extended Poincar\'e symmetry group are explained in Section~\ref{section4}.  In this very section we also explore the relativistic features of DFRA space and construct the Klein--Gordon equation together with the Noether's formalism.  In Section~\ref{section5} we analyzed the issue about the fermions in this new structure and we study the Dirac equation.  Finally, in Section~\ref{section6} we complete the review considering the elaboration of complex scalar f\/ields and the source term analysis of quantum f\/ield theories.

\section{The noncommutative  quantum mechanics}\label{section2}

In this section we will introduce the complete extension of the DFR space formulated by R.~Amorim \cite{Amorim1,Amorim4,Amorim5,Amorim2,aa} where a minimal extension of the DFR is accomplished through the introduction of the canonical conjugate momenta to the variable $\theta^{\mu\nu}$ of the system.

In the last section we talked about the DFR space, its physical motivations and main mathematical ingredients.  Concerning now the DFRA space, we continue to furnish its ``missing parts'' and naturally its implications in quantum mechanics.

\subsection{The Doplicher--Fredenhagen--Roberts--Amorim space}

The results appearing in \cite{Carlson} motivated some other authors
 \cite{Haghighat,Carone,Ettefaghi,Morita,Saxell}. Some of them prefer  to start from the beginning  by adopting the Doplicher, Fredenhagen and Roberts (DFR) algebra \cite{DFR}, which essentially assumes
(\ref{a1}) as well as the vanishing of the triple commutator among the coordinate operators,
\begin{gather} \label{triple}
[{\mathbf x}^{\mu},[{\mathbf x}^{\nu},{\mathbf x}^{\rho}]]=0,
\end{gather}
and it is easy to realize that this relation constitute a constraint in a NC spacetime.  Notice that the commutator inside the triple one is not a $c$-number.

The DFR algebra is based on  principles imported from general relativity (GR) and quantum mechanics (QM). In addition to (\ref{a1}) it also assumes that
\begin{gather}
\label{6a}
[{\mathbf x}^\mu, \theta^{\alpha\beta}] = 0 ,
\end{gather}
and we consider that space has arbitrary $D\geq2$ dimensions.  As usual ${\mathbf x}^\mu$ and ${\mathbf p}_\nu$, where $i,j=1,2,\dots,D$ and $\mu,\nu=0,1,\dots,D$,  represent the position operator and its
conjugate momentum.   The NC variable $\theta^{\mu\nu}$ represent the noncommutativity operator, but now $\pi_{\mu\nu}$ is its conjugate momentum. In accordance with the discussion above, it follows the algebra
\begin{subequations}\label{7000}
\begin{gather}
[{\mathbf x}^\mu,{\mathbf p}_\nu] = i \delta^\mu_\nu ,\label{7000a}\\
[\theta^{\mu\nu},\pi_{\alpha\beta}] = i \delta^{\mu\nu}_{\,\,\,\,\alpha\beta},\label{7000b}
\end{gather}
\end{subequations}
 where $\delta^{\mu\nu}_{\,\,\,\,\alpha\beta}=\delta^{\mu}_{\alpha}\delta^{\nu}_{\beta}-\delta^{\mu}_{\beta}\delta^{\nu}_{\alpha}$.
The relation (\ref{a1}) here in a space with $D$ dimensions, for example, can be written as
\begin{gather}
\label{9000}
[{\mathbf x}^i,{\mathbf x}^j] = i  \theta^{ij} \qquad \mbox{and} \qquad [{\mathbf p}_i,{\mathbf p}_j ] = 0
\end{gather}
and together with the triple commutator (\ref{triple}) condition of the standard spacetime, i.e.,
\begin{gather}
\label{10}
[{\mathbf x}^\mu,\theta^{\nu\alpha}] = 0.
\end{gather}
This implies that
\begin{gather}
\label{11}
[\theta^{\mu\nu},\theta^{\alpha\beta}] = 0 ,
\end{gather}
and this completes the DFR algebra.

Recently, in order to obtain consistency R.~Amorim introduced~\cite{Amorim1}, as we talked above, the canonical conjugate momenta $\pi_{\mu\nu}$ such that,
\begin{gather}
\label{12}
[{\mathbf p}_\mu,\theta^{\nu\alpha}] = 0,\qquad
[{\mathbf p}_\mu,\pi_{\nu\alpha}] = 0.
\end{gather}

The Jacobi identity formed by the operators ${\mathbf x}^i$, ${\mathbf x}^j$ and $\pi_{kl}$ leads to the nontrivial relation
\begin{gather}
\label{14}
[[{\mathbf x}^\mu,\pi_{\alpha\beta}],{\mathbf x}^\nu]- [[{\mathbf x}^\nu,\pi_{\alpha\beta}],{\mathbf x}^\mu]   =   - \delta^{\mu\nu}_{\,\,\,\,\alpha\beta}.
\end{gather}
The solution, unless trivial terms,  is given by
\begin{gather}
\label{15000}
[{\mathbf x}^\mu,\pi_{\alpha\beta}]=-{i\over 2}\delta^{\mu\nu}_{\,\,\,\,\alpha\beta}{\mathbf p}_\nu.
\end{gather}
It is simple to verify that the whole set of commutation relations listed above is indeed consistent under all possible Jacobi identities. Expression (\ref{15000}) suggests the  shifted coordinate
operator \cite{Chaichan,Gamboa,Kokado,Kijanka,Calmet}
\begin{gather}
\label{16000}
{\mathbf X}^\mu\equiv{\mathbf x}^\mu + {1\over 2}\theta^{\mu\nu}{\mathbf p}_\nu
\end{gather}
that commutes with $\pi_{kl}$. Actually, (\ref{16000}) also commutes with $\theta^{kl}$ and $ {\mathbf X}^j $, and satisf\/ies a non trivial commutation relation with  ${\mathbf p}_i$  depending objects, which could be derived from
\begin{gather}
\label{17000}
[{\mathbf X}^\mu,{\mathbf p}_\nu]=i\delta^\mu_\nu
\end{gather}
and
\begin{gather}
\label{i6}
[{\mathbf X}^\mu,{\mathbf X}^\nu]=0 .
\end{gather}

To construct a DFRA algebra in $(x,\theta)$ space, we can write
\[
{\mathbf M}^{\mu\nu} = {\mathbf X}^\mu{\mathbf p}^\nu - {\mathbf X}^\nu{\mathbf p}^\mu - \theta^{\mu\sigma} \pi_{\sigma}^{\:\:\nu} + \theta^{\nu\sigma} \pi_{\sigma}^{\:\:\mu} ,
 \]
 where ${\mathbf M}^{\mu\nu}$ is the antisymmetric generator of the Lorentz-group.  To construct  $\pi_{\mu\nu}$ we have to obey equations
(\ref{7000b}) and (\ref{15000}), obviously.   From (\ref{7000a}) we can write the generators of translations as
\[
P_\mu = - i \partial_\mu.
\]
  With these ingredients it is easy to construct the commutation relations
\begin{gather*}
\left[ {\mathbf P}_\mu , {\mathbf P}_\nu \right]  =  0, \qquad
\left[ {\mathbf M}_{\mu\nu},{\mathbf P}_{\rho} \right]  =  - i \big(\eta_{\mu\nu} {\mathbf P}_\rho - \eta_{\mu\rho} {\mathbf P}_\nu \big),\\ % \\\label{ABC} \\
\left[ {\mathbf M}_{\mu\nu},{\mathbf M}_{\rho\sigma} \right]  =  - i \big( \eta_{\mu\rho} {\mathbf M}_{\nu\sigma} - \eta_{\mu\sigma} {\mathbf M}_{\nu\rho} - \eta_{\nu\rho} {\mathbf M}_{\mu\sigma} - \eta_{\nu\sigma} {\mathbf M}_{\mu\rho} ) , \nonumber
\end{gather*}
and we can say that ${\mathbf P}_\mu$ and ${\mathbf M}_{\mu\nu}$ are the generator of the DFRA algebra.  These relations are important, as we will see in Section~\ref{section5}, because they are essential for the extension of the Dirac equation to the extended DFRA conf\/iguration space $(x,\theta)$.  It can be shown that  the Clif\/ford algebra structure generated by the 10 generalized Dirac matrices $\Gamma$ (Section~\ref{section5}) relies on these relations.

Now we need to remember some basics in quantum mechanics.  In order to introduce a~continuous basis for a~general  Hilbert space, with the aid of the above commutation relations, it is necessary f\/irstly to f\/ind a maximal set of commuting operators. For instance, let us choose a momentum basis formed by the eigenvectors of ${\mathbf p}$ and
$\pi$. A coordinate basis formed by the eigenvectors of (${\mathbf X},\theta$) can also be introduced, among other possibilities. We observe here that  it is in no way possible to form a basis involving more than one component of the original position operator~${\mathbf x}$, since their components do not commute.

To clarify, let us display the fundamental relations involving those basis, namely  eigenvalue, orthogonality and completeness relations
\begin{gather}
%\label{18}
{\mathbf X}^i |{ X}',{ \theta}'\rangle = {X'}^i|{ X}',{ \theta}'\rangle
 ,\qquad {\mathbf\theta}^{ij} |{ X}',{ \theta}'\rangle = {\theta'}^{ij}|{ X}',{ \theta}'\rangle  ,
\nonumber\\
%\label{19}
{\mathbf p}_i |{ p}',{ \pi}'\rangle = {p'}_i|{ p}',{ \pi}'\rangle ,\qquad {\mathbf\pi}_{ij} |{ p}',{ \pi}'\rangle = {\pi'}_{ij}|{ p}',{ \pi}'\rangle ,
\nonumber\\
%\label{20}
\langle  { X}',{ \theta}'|{ X}'',{ \theta}''\rangle = \delta^D (X'-X'')\delta^{\frac{D(D-1)}{2}}({\theta}'-{\theta}'') ,
\nonumber\\
%\label{21}
\langle { p}',{ \pi}'|{ p}'',{ \pi}''\rangle = \delta^D(p'-p'')\delta^{\frac{D(D-1)}{2}}({\pi}'-{\pi}'') ,
\nonumber\\
\label{22}
\int d^D X'\,d^{\frac{D(D-1)}{2}}{\theta'}  |{ X}',{ \theta}'\rangle \langle { X}',{ \theta}'|= {\mathbf 1},
\\
\label{23}
\int d^D p'\,d^{\frac{D(D-1)}{2}}{\pi'}  |{ p}',{ \pi}'\rangle \langle { p}',{ \pi}'|= {\mathbf 1} ,
\end{gather}
notice that the dimension $D$ means that we live in a framework formed by the spatial coordinates and by the $\theta$ coordinates, namely, $D$~includes both spaces, $D=$ (spatial coordinates $+ \theta$ coordinates).  It can be seen clearly from the equations involving the delta functions and the integrals equations~(\ref{22}) and~(\ref{23}).

Representations of the operators in those bases can be obtained in an usual way. For instance, the commutation relations given by equations~(\ref{7000}) to~(\ref{17000}) and the eigenvalue relations above, unless trivial terms, give
\begin{gather*}
%\label{24}
\langle  { X}',{ \theta}'|{\mathbf p}_i|{ X}'',{ \theta}''\rangle = -i{\frac{\partial}{\partial X'^i}}\delta^D (X'-X'')\delta^{\frac{D(D-1)}{2}}({\theta}'-{\theta}'')
\end{gather*}
and
\begin{gather*}
%\label{25}
\langle  { X}',{ \theta}'|\pi_{ij}|{ X}'',{ \theta}''\rangle
 =  -i\delta^D (X'-X''){\frac{\partial}{\partial \theta'^{ij}}}\delta^{\frac{D(D-1)}{2}}({\theta}'-{\theta}'') .
\end{gather*}
The transformations from one basis to the other one are carried out by extended Fourier transforms. Related with these transformations is the ``plane wave''
\begin{gather*} %\label{25.1}
\langle  { X}',{ \theta}'|{ p}'',{ \pi}''\rangle = N \exp ( i p'' {X'}+  i{\pi''}{\theta'}) ,
\end{gather*}
where internal products are represented in a compact manner. For instance,
\begin{gather*} %\label{25.2}
p'' {X'}+  {\pi''}{\theta'}=  p''_i {X'}^i+ {\frac{1}{2}} \pi''_{ij}{\theta'}^{ij} .
\end{gather*}

Before discussing any dynamics, it seems interesting to study the generators of the group of rotations $SO(D)$. Without considering the spin sector, we realize that the usual angular momentum operator
\begin{gather*}
%\label{27}
{\mathbf l}^{ij}= {\mathbf x}^i{\mathbf p}^j-{\mathbf x}^j{\mathbf p}^i
\end{gather*}
does not close in an algebra due to (\ref{9000}). And we have that,
\begin{gather*}
 [{\mathbf l}^{ij},{\mathbf l}^{kl}]=i\delta^{il}{\mathbf l}^{kj}-i\delta^{jl}{\mathbf l}^{ki}-i\delta^{ik}{\mathbf l}^{lj}+i\delta^{jk}{\mathbf l}^{li}
-i\theta^{il}{\mathbf p}^{k}{\mathbf p}^{j}+i\theta^{jl}{\mathbf p}^{k}{\mathbf p}^{i}+i\theta^{ik}{\mathbf p}^{l}
{\mathbf p}^{j}-i\theta^{jk}{\mathbf p}^{l}{\mathbf p}^{i}  %\label{28}
\end{gather*}
and so their components can not be  $SO(D)$ generators in this extended Hilbert space.
On the contrary, the  operator
\begin{gather}
\label{29}
{\mathbf L}^{ij}= {\mathbf X}^i{\mathbf p}^j-{\mathbf X}^j{\mathbf p}^i ,
\end{gather}
closes in the $SO(D)$ algebra. However, to properly  act  in the ($\theta,\pi$) sector, it has to be generalized  to the total angular momentum operator
\begin{gather}
\label{30}
{\mathbf J}^{ij}= {\mathbf L}^{ij}-\theta^{il}\pi_l^{\,\,j}+\theta^{jl}\pi_l^{\,\,i} .
\end{gather}
It is easy to see that not only
\begin{gather}
\label{31}
[{\mathbf J}^{ij},{\mathbf J}^{kl}]=i\delta^{il}{\mathbf J}^{kj}-i\delta^{jl}{\mathbf J}^{ki}-i\delta^{ik}{\mathbf J}^{lj}+i\delta^{jk}{\mathbf J}^{li} ,
\end{gather}
but ${\mathbf J}^{ij}$ generates  rotations in all Hilbert space sectors. Actually
\begin{alignat}{3}
& \delta {\mathbf X}^i={i\over2}\epsilon_{kl}[ {\mathbf X}^i, {\mathbf J}^{kl}] =\epsilon^{ik}{\mathbf X}_k, \qquad&&
\delta {\mathbf p}^i={i\over2}\epsilon_{kl}[ {\mathbf p}^i, {\mathbf J}^{kl}]=\epsilon^{ik}{\mathbf p}_k,&\nonumber \\
& \delta \theta^{ij}={i\over2}\epsilon_{kl}[ \theta^{ij}, {\mathbf J}^{kl}]=\epsilon^{ik}\theta_k^{\,\,j}+
\epsilon^{jk}\theta^i_{\,\,k},\qquad &&
\delta \pi^{ij}={i\over2}\epsilon_{kl}[ \pi^{ij}, {\mathbf J}^{kl}] =\epsilon^{ik}\pi_k^{\,\,j}+
\epsilon^{jk}\pi^i_{\,\,k} & \label{32000}
\end{alignat}
have the expected form. The same occurs with
\begin{gather*} %\label{32.1}
{\mathbf x}^i={\mathbf X}^i-{1\over 2}\theta^{ij}{\mathbf p}_j\quad\Longrightarrow\quad
\delta {\mathbf x}^i={i\over2}\epsilon_{kl} [ {\mathbf x}^i, {\mathbf J}^{kl}] =\epsilon^{ik}{\mathbf x}_k .
\end{gather*}
Observe that in the usual NCQM prescription, where the objects of noncommutativity are parameters or where the angular momentum operator has not been generalized, ${\mathbf X}$ fails to transform as a vector operator under $SO(D)$ \cite{Chaichan,Gamboa,Kokado,Kijanka,Calmet}. The consistence of transformations (\ref{32000}) comes from the fact that they are generated through the action of a symmetry operator and not from operations based on the index structure of those variables.

We would like to mention that in $D=2$ the operator ${\mathbf J}^{ij}$ reduces to ${\mathbf L}^{ij}$, in accordance with the fact that in this case $\theta$ or $\pi$ has only one independent component. In $D=3$, it is possible to represent $\theta$ or $ \pi$ by three vectors and both parts of the angular momentum operator have the same kind of structure, and so the same spectrum. An unexpected addition of angular momentum potentially arises, although the ($\theta,\pi$) sector can live in a ${\mathbf J}=0$ Hilbert subspace.
Unitary  rotations are generated by $U(\omega)=\exp(-i\omega \cdot {\mathbf J})$, while unitary translations, by $T(\lambda,\Xi)=\exp(-i\lambda \cdot {\mathbf p}-i\Xi \cdot \pi) $.

\subsection{The noncommutative harmonic oscillator}

In this section we will consider the isotropic $D$-dimensional harmonic oscillator where we f\/ind several possibilities of rotational invariant Hamiltonians which  present the proper commutative limit \cite{Gamboa,Nair,Kijanka,Dadic}.
The well known expression representing the harmonic oscillator can be written as
\begin{gather}
\label{34000}
{\mathbf H}_0={\frac{1}{2m}}{\mathbf p}^2+{\frac{m\omega^2}{2}}{\mathbf X}^2 ,
\end{gather}
since ${\mathbf X}^i$ commutes with ${\mathbf X}^j$, satisf\/ies the canonical relation
(\ref{17000}) and in the DFRA formalism transforms according to (\ref{32000}).
With these results we can construct   annihilation and creation operators  in the usual way,
\begin{gather*} %\label{34.1}
{\mathbf A}^i=\sqrt{{\frac{m\omega}{2}}}\left({\mathbf X}^i+{\frac{i{\mathbf p}^i}{m\omega}}\right) \qquad \mbox{and} \qquad
{\mathbf A}^{\dag i}=\sqrt{{\frac{m\omega}{2}}}\left({\mathbf X}^i-{\frac{i{\mathbf p}^i}{m\omega}}\right) ,
\end{gather*}
where ${\mathbf A}^i$ and ${\mathbf A}^{\dagger i}$ satisfy the usual harmonic oscillator algebra, and   ${\mathbf H}_0$ can be written in terms of the sum  of $D$ number operators      ${\mathbf N}^i={\mathbf A}^{\dag i}{\mathbf A}^i$, which have  the same spectrum and the same degeneracies when compared with the ordinary QM case~\cite{Cohen}.

The ($\theta,\pi$) sector, however, is not modif\/ied by any new dynamics
if ${\mathbf H}_0$ represents the total Hamiltonian.
As the harmonic oscillator describes a system near an equilibrium  conf\/iguration, it seems interesting as well to add to (\ref{34000}) a new term like
\begin{gather}
\label{35}
{\mathbf H}_\theta=
{\frac{1}{2\Lambda}}\pi^2+{\frac{\Lambda\Omega^2}{2}}\theta^2 ,
\end{gather}
where $\Lambda$ is a parameter with dimension of $({\rm length})^{-3}$ and $\Omega$ is some frequency.
Both Hamiltonians, equations (\ref{34000}) and (\ref{35}), can be simultaneously diagonalized, since they commute.
Hence, the total Hamiltonian eigenstates will be formed by the direct product of the Hamiltonian eigenstates of each sector.

The annihilation  and creation operators, considering the ($\theta,\pi$) sector, are respectively def\/ined as
\begin{gather*} %\label{35.1}
{\mathbf A}^{ij}=\sqrt{{\frac{\Lambda\Omega}{2}}}\left(\theta^{ij}+{\frac{i\pi^{ij}}{\Lambda\Omega}}\right) \qquad \mbox{and} \qquad
{\mathbf A}^{\dag  ij}=\sqrt{{\frac{\Lambda\Omega}{2}}}\left(\theta^{ij}-{\frac{i\pi^{ij}}{\Lambda\Omega}}\right),
\end{gather*}
which satisfy the oscillator algebra
\begin{gather*}
[{\mathbf A}^{ij},{\mathbf A}^{\dag  kl}]=\delta^{ij,kl} ,
\end{gather*}
and now we can construct eigenstates of $H_\theta$, equation \eqref{35}, associated with  quantum numbers~$n^{ij}$. As well known, the ground state is annihilated by ${\mathbf A}^{ij}$, and its  corresponding wave function,
in the ($\theta,\pi$) sector, is
\begin{gather}
\label{38}
\langle \theta'|n^{ij}=0,t\rangle =\left({\frac{\Lambda\Omega}{\pi}}\right)^{\frac{D(D-1)}{8}}\exp\left[-{\frac{\Lambda\Omega}{4}}{\theta'}_{ij}{\theta'}^{ij}\right]
\exp\left[-iD(D-1){\Omega\over 4} t\right].
\end{gather}
 However, turning to the basics, the wave functions for excited states can be obtained through the application of the creation operator ${\mathbf A}^{\dag kl}$ on the fundamental state. On the other hand, we expect  that~$\Omega$ might be so big that  only the fundamental level of this generalized oscillator could  be occupied. This will generate only a shift in the oscillator spectrum, which is $\Delta E={\frac{D(D-1)}{4}}\Omega$ and
this new vacuum energy could generate unexpected behaviors.

Another point related with (\ref{38}) is that it gives  a natural way for introducing  the weight
function $W(\theta)$ which appears, in the context of NCFT's, in~\cite{Carlson, Morita}.  $W(\theta)$ is a normalized function necessary, for example, to control the $\theta$-integration.
Analyzing the ($\theta,\pi$) sector, the expectation value of any  function  $f(\theta )$ over the fundamental state is
\begin{gather*}
\langle f(\theta)\rangle =\langle n^{kl}=0,t|f(\theta)|n^{kl}=0,t\rangle \nonumber\\
\phantom{\langle f(\theta)\rangle}{}
=\left({\frac{\Lambda\Omega}{\pi}}\right)^{{\frac{D(D-1)}{4}}}
\int  d^{\frac{D(D-1)}{2}}{\theta'}   f(\theta')\exp\left[-{\frac{\Lambda\Omega}{2}}{\theta'}_{rs}{\theta'}^{rs}\right]
\equiv \int d^{\frac{D(D-1)}{2}}{\theta'}  W({\theta'})  f(\theta'),%\label{39}
\end{gather*}
where
\begin{gather*}
%\label{40}
W({\theta'})\equiv\left({\frac{\Lambda\Omega}{\pi}}\right)^{{\frac{D(D-1)}{4}}}
\exp\left[-{\frac{\Lambda\Omega}{2}}{\theta'}_{rs}{\theta'}^{rs}\right],
\end{gather*}
and the expectation values are given by
\begin{gather}
\langle {\mathbf 1}\rangle =1,\qquad
\langle \theta^{ij}\rangle = 0,\qquad
{1\over2}\langle \theta^{ij}\theta_{ij}\rangle =\langle \theta^2\rangle,\qquad
\langle \theta^{ij}\theta^{kl}\rangle = {\frac{2}{D(D-1)}}\delta^{ij,kl}\langle \theta^2\rangle,\label{41}
\end{gather}
where
\[
\langle \theta^2\rangle \equiv {\frac{1}{2\Lambda \Omega}}
\]
and now we can calculate the expectation values for the physical coordinate operators. As one can verify,
$\langle {\mathbf x}^i\rangle =\langle {\mathbf X}^i\rangle =0$, but one can f\/ind non trivial noncommutativity contributions to the expectation values for other operators. For instance, it is easy to see from~(\ref{41}) and~(\ref{16000}) that
\[
\langle {\mathbf x}^2\rangle =\langle {\mathbf X}^2\rangle + {2\over D}\langle \theta^2\rangle \langle {\mathbf p}^2\rangle ,
\]
 where $\langle {\mathbf X}^2\rangle $ and $\langle {\mathbf p}^2\rangle$ are the usual QM results for an isotropic oscillator
in a given state. This shows that noncommutativity enlarges the root-mean-square deviation for the physical coordinate operator, as expected and can be measurable, at f\/irst sight.

\section{Tensor coordinates in noncommutative  mechanics}\label{section3}

All the operators introduced until now belong to the same algebra and are equal, hierarchically speaking.  The necessity of a rotation invariance under the group ${\rm SO}(D)$ is a consequence of this augmented Hilbert space.  Rotation invariance, in a nonrelativistic theory, is the main topic if one  intends to describe any physical system in a consistent way.

In NCFT's it is possible to achieve the corresponding $SO(D,1)$  invariance also by promo\-ting~$\theta^{\mu\nu}$ from a constant matrix to a tensor operator~\cite{Carlson, Banerjee2,Morita,Haghighat,Carone,Ettefaghi,Saxell}, although in this last situation the rules are quite dif\/ferent from those found in NCQM, since in a quantum f\/ield theory  the relevant operators are not coordinates but f\/ields.

Now that we got acquainted with the new proposed version of NCQM \cite{Amorim1} where the $\theta^{ij}$ are tensors in Hilbert space and $\pi_{ij}$ are their conjugate canonical momenta, we will show in this section that a possible fundamental classical theory, under quantization, can reproduce the algebraic structure depicted in the last section.

The Dirac formalism \cite{Dirac} for constrained Hamiltonian systems is extensively used for this purpose.
As it is well known, when a theory presents a complete set of second-class constraints $\Xi^a=0$, $a=1,2,\ldots,2N$, the Poisson brackets  $\{A,B\}$ between any two phase space quantities~$A$,~$B$ must be replaced by Dirac brackets{\samepage
\begin{gather}
\label{15001}
\{A,B\}_D=\{A,B\}-\{A,\Xi^a\}\Delta^{-1}_{ab}\{\Xi^b,B\} ,
\end{gather}
such that the evolution of the system respects the constraint surface given by   $\Xi^a=0$.}

In (\ref{15001})
\begin{gather}
\label{16001}
\Delta^{ab}=\{\Xi^a,\Xi^b\}
\end{gather}
is  the so-called constraint matrix and $\Delta^{-1}_{ab}$ is its inverse.
The fact that the constraints $\Xi^a$ are second-class guarantees the existence of $\Delta^{ab}$.
If that matrix were singular,  linear combinations of the $\Xi^a$ could  be f\/irst class. For the f\/irst situation, the number of ef\/fective degrees of freedom
of the theory is given by $2\cal{D}$ $-2N$, where $2\cal{D}$ is the number of phase space variables and $2N$ is the number of second-class constraints.

If the phase space is described only by the
\[
2{\cal{D}}=2D+2{\frac{D(D-1)}{2}}
\]
variables $x^i$, $p_i$, $\theta^{ij}$ and $\pi_{ij}$, the introduction of second-class constraints  generates an over constrained theory when compared with the algebraic structure given in the last section.
Consequently, it seems necessary to enlarge the phase space by $2N$ variables, and to introduce at the same time $2N$ second-class constraints.
An easy way to implement these concepts without destroying the symmetry under rotations is to enlarge the phase space introducing a pair of canonical variables $Z^i$, $K_i$, also with (at the same time) a set of  second-class constraints $\Psi^i$, $\Phi_i$.

Considering this set of phase space variables, it follows by construction the  fundamental (non vanishing) Poisson bracket structure
\begin{gather}
\label{17001}
\{x^i,p_j\} = \delta^i_j,
\qquad
\{\theta^{ij},\pi_{kl}\} = \delta^{ij}_{\,\,\,\,kl},
\qquad
\{Z^i,K_j\} = \delta^i_j
\end{gather}
and the Dirac brackets structure is derived in accordance with the form of the second-class constraints, subject that will be discussed in what follows.

Let us assume that  $Z^i$ has  dimension of length $L$, as $x^i$. This implies that both $p_i$ and $K_i$ have dimension of $L^{-1}$.
As $\theta^{ij}$ and $\pi_{ij}$ have dimensions of $L^2$ and $L^{-2}$ respectively,  the expression for the  constraints $\Psi^i$ and $\Phi_i$ is given by
\[
\Psi^i=Z^i+\alpha x^i+\beta\theta^{ij}p_j+\gamma\theta^{ij}K_j
\]
and
\[
\Phi_i=K_i+\rho p_i+\sigma\pi_{ij}x^j+\lambda\pi_{ij}Z^j,
\]
if only dimensionless parameters $\alpha$, $\beta$, $\gamma$, $\rho$, $\sigma$ and $\lambda$ are introduced and any power higher than two in phase space variables is  discarded. It is possible to display the whole group of parameters during the computation of the Dirac formalism.  After that, at the end of the calculations, the parameters have been chosen in order to generate, under quantization,  the commutator structure appearing in equations (\ref{7000}) to (\ref{15000}).  The constraints  reduce, in this situation, to
\begin{gather}
\label{18001}
\Psi^i = Z^i-{1\over2}\theta^{ij}p_j,
\qquad
\Phi_i = K_i-p_i
\end{gather}
and hence the corresponding constraint matrix (\ref{16001}) becomes
\begin{gather}
\label{19001}
(\Delta^{ab})= \left( \begin{array}{cc}
        \{\Psi^i,\Psi^j\}&\{\Psi^i,\Phi_j\}\\
	\{\Phi_i,\Psi^j\}&\{\Phi_i,\Phi_j\}
\end{array} \right)
=
\left( \begin{array}{cc}
         0 & \delta^i_{j}\\
	-\delta^{j}_i & 0
\end{array} \right) .
\end{gather}
Notice that (\ref{19001}) is regular even if $\theta^{ij}$ is singular. This fact guarantees that the proper commutative limit of the theory can be taken.

The inverse of (\ref{19001}) is trivially given by
\begin{gather*}
%\label{20001}
(\Delta^{-1}_{ab})= \left( \begin{array}{cc}
        0&-\delta^{\,\,j}_i \\
	\delta^{i}_{\,\,j}&0 \end{array} \right)
\end{gather*}
and it is easy to see that the  non-zero Dirac brackets (the others are zero) involving only the original set of phase space variables are
\begin{gather}
\{x^i,p_j\}_D=\delta^i_j, \qquad \{x^i,x^j\}_D=\theta^{ij},\qquad
\{\theta^{ij},\pi_{kl}\}_D=\delta^{ij}_{\,\,\,kl}, \nonumber\\
\{x^i,\pi_{kl}\}_D=-{1\over2}\delta^{ij}_{\,\,\,kl}\,p_j ,\label{21001}
\end{gather}
which furnish the desired result. If $y^A$ represents phase space variables and ${\mathbf y}^A$ the corresponding Hilbert space operators, the Dirac quantization procedure,
\[
\{y^A,y^B\}_D\rightarrow {1\over i}[{\mathbf y}^A,{\mathbf y}^B]
\]
results the commutators in (\ref{7000}) until (\ref{15000}).  For completeness, the remaining non-zero Dirac brackets involving $Z^i$ and $K_i$ are
\begin{gather}
\label{22001}
\{Z^i,x^j\}_D=-{1\over2}\theta^{ij},\qquad
\{K_i,x^j\}_D=-\delta^j_i,\qquad
\{Z^i,\pi_{kl}\}_D={1\over2}\delta^{ij}_{\,\,\,kl}p_j .
\end{gather}

In this classical theory the shifted coordinate
\begin{gather*}
%\label{24001}
X^i=x^i+{1\over2}\theta^{ij}p_j,
\end{gather*}
which corresponds  to the operator (\ref{16000}), also plays a fundamental role. As can be verif\/ied by the non-zero Dirac brackets just below,
\begin{gather*}
\{X^i,p_j\}_D=\delta^i_j,\qquad
\{X^i,x^j\}_D={1\over2} \theta^{ij},\qquad
\{X^i,Z^j\}_D=-{1\over2} \theta^{ij},\qquad
        \{X^i,K_j\}_D=\delta^i_j ,%\label{25001}
\end{gather*}
and the angular momentum tensor
\begin{gather}
\label{26001}
{ J}^{ij}= { X}^i{ p}^j-{ X}^j{ p}^i-{ \theta}^{il}{ \pi}_l^{\,\,j}+{ \theta}^{jl}{ \pi}_l^{\,\,i}
\end{gather}
closes in the classical $SO(D)$ algebra, by using Dirac brackets instead of commutators. In fact
\begin{gather*}
%\label{27001}
\{{ J}^{ij},{ J}^{kl}\}_D=\delta^{il}{ J}^{kj}-\delta^{jl}{ J}^{ki}-\delta^{ik}{ J}^{lj}+\delta^{jk}{ J}^{li} ,
\end{gather*}
and as in the quantum case,  the proper symmetry transformations over all the phase space variables are generated by
(\ref{26001}).  Beginning with
\begin{gather*}
%\label{28001}
\delta A=-{1\over2}\epsilon_{kl}\{A,{ J}^{kl}\}_D ,
\end{gather*}
one have as a result that
\begin{gather*}
\delta X^i=\epsilon ^i_{\,\,j} X^j,\qquad
\delta x^i=\epsilon ^i_{\,\,j} x^j,\qquad
\delta p_i=\epsilon _i^{\,\,j} p_j,\qquad
\delta \theta^{ij}=\epsilon ^i_{\,\,k} \theta^{kj}+ \epsilon ^j_{\,\,k} \theta^{ik},\nonumber\\
\delta \pi_{ij}=\epsilon _i^{\,\,k} \pi_{kj}+ \epsilon _j^{\,\,k} \pi_{ik},\qquad
\delta Z^i={1\over2}\epsilon ^i_{\,\,j} \theta^{jk}p_k,\qquad
\delta K_i=\epsilon _i^{\,\,\,j} p_j .%\label{29001}
\end{gather*}
The last two equations above also furnish the proper result on the constraint surface. Hence, it was possible to generate all the desired structure displayed in the last section by using the Dirac brackets and the constraints given in~(\ref{18001}). These constraints, as well as the fundamental Poisson brackets in~(\ref{17001}), can be easily generated by the f\/irst order action
\begin{gather}
\label{30001}
S=\int dt \, L_{\rm FO},
\end{gather}
where
\begin{gather}
\label{31001}
L_{\rm FO}=p \cdot \dot x + K\cdot \dot Z+\pi\cdot \dot\theta - \lambda_a\Xi^a-H .
\end{gather}
The  $2D$ quantities $\lambda_a$ are the Lagrange multipliers introduced conveniently to implement the constraints $\Xi^a=0$ given by~(\ref{18001}), and $H$ is some Hamiltonian. The dots ``$\cdot$'' between phase space coordinates represent internal products.
The canonical conjugate momenta for the Lagrange multipliers are primary constraints that, when conserved, generate the secondary constraints $\Xi^a=0$. Since these  last constraints are second class, they are automatically conserved by the theory, and the Lagrange multipliers are determined in the process.

The general expression  for the f\/irst-order Lagrangian in (\ref{31001}) shows the constraints implementation in this enlarged space, which is a trivial result, analogous to the standard procedure through the Lagrange multipliers.  To obtain a more illuminating second-order Lagrangian we must follow the basic pattern and with the help of the Hamiltonian, integrate out the momentum variables in~(\ref{31001}).

As we explained in the last section, besides the introduction of the referred algebraic structure, a specif\/ic Hamiltonian has been furnished, representing a generalized isotropic harmonic oscillator, which contemplates with dynamics not only   the usual vectorial coordinates but also the noncommutativity sector spanned by the tensor quantities $\theta$ and $\pi$. The corresponding classical Hamiltonian can be written as
\begin{gather}
\label{32001}
{ H}={\frac{1}{2m}}{ p}^2+{\frac{m\omega^2}{2}}{ X}^2+
{\frac{1}{2\Lambda}}{ \pi}^2+{\frac{\Lambda\Omega^2}{2}}{ \theta}^2 ,
\end{gather}
which is invariant under rotations. In (\ref{32001}) $m$ is a mass, $\Lambda$ is a parameter with dimension of $L^{-3}$, and~$\omega$ and~$\Omega$ are frequencies.  Other choices for the Hamiltonian  can be done without spoiling the algebraic structure discussed above.

The classical system given by (\ref{30001}), (\ref{31001}) and (\ref{32001}) represents two independent isotropic oscillators in~$D$ and ${\frac{D(D-1)}{2}}$ dimensions, expressed in terms of variables $X^i$, $p_i$, $\theta^{ij}$ and~$\pi_{ij}$. The solution is elementary, but when one expresses the oscillators in terms of physical variables $x^i$, $p_i$, $\theta^{ij}$ and $\pi_{ij}$, an interaction appears between them, with cumbersome equations of motion. In this sense the former set of variables gives, in the phase space,  the  normal coordinates that decouple both oscillators.

It was possible to generate a Dirac brackets algebraic structure that, when quantized, reproduce exactly the commutator algebra appearing in the last section. The presented  theory has been proved to be invariant under the action of the rotation group ${\rm SO}(D)$   and could be derived through a variational principle.
Once this structure has been given, it is not dif\/f\/icult to construct a relativistic generalization of such a model. The fundamental Poisson brackets become
\begin{gather*}
%\label{33001}
\{x^\mu,p_\nu\} = \delta^\mu_\nu,
\qquad
\{\theta^{\mu\nu},\pi_{\rho\sigma}\} = \delta^{\mu\nu}_{\,\,\,\,\rho\sigma},\qquad
\{Z^\mu,K_\nu\} = \delta^\mu_\nu ,
\end{gather*}
and the constraints (\ref{18001}) are generalized to
\begin{gather*}
%\label{34001}
\Psi^\mu = Z^\mu-{1\over2}\theta^{\mu\nu}p_\nu,
\qquad
\Phi_\mu = K_\mu-p_\mu ,
\end{gather*}
generating the invertible constraint matrix
\begin{gather*}
%\label{35001}
(\Delta^{ab})= \left( \begin{array}{cc}
        \{\Psi^\mu,\Psi^\nu \} & \{\Psi^\mu,\Phi^\nu \} \\
	\{\Phi^\mu,\Psi^\nu \} & \{\Phi^\mu,\Phi^\nu \} \end{array} \right)
= \left( \begin{array}{cc}
          0 & \eta^{\mu\nu} \\
	-\eta^{\mu\nu} & 0 \end{array} \right) .
\end{gather*}
To f\/inish we can say that the Dirac brackets between the phase space variables can also be generalized from
(\ref{21001}), (\ref{22001}). The Hamiltonian of course cannot be given by~(\ref{32001}), but at least for the free particle, it vanishes identically, as it is usual to appear with covariant classical systems~\cite{Dirac}. Also it is necessary for a new constraint, which must be f\/irst class, to generate the  reparametrization transformations. In a minimal extension of the usual commutative case, it is given by the mass shell condition
\[
\chi=p^2+m^2=0,
\]
but other choices are possible, furnishing dynamics to the noncommutativity sector or enlarging the symmetry content of the relativistic action.

\section{Dynamical symmetries in NC theories}\label{section4}

In this section we will analyze the dynamical spacetime symmetries   in NC relativistic theories by using the DFRA algebra depicted in Section~\ref{section2}. As explained there, the formalism is constructed in an extended spacetime with independent degrees of freedom associated with the object of noncommutativity $\theta^{\mu\nu}$. In this framework we can consider theories that are invariant under the Poincar\'{e}   group ${\cal P}$ or under its extension ${\cal P}'$, when translations in the extra dimensions are allowed. The Noether formalism adapted to such extended $x+\theta$ spacetime will be employed.

We will study the algebraic structure of the generalized coordinate operators and their conjugate momenta, and construct the appropriate representations for the generators of ${\cal P}$ and ${\cal P}'$, as well as for the associated Casimir operators.
Next, some possible NCQM actions constructed with those Casimir operators will be introduced and after that we
will investigate the symmetry content of one of those theories by using Noether's procedure.

\subsection{Coordinate operators  and their transformations in relativistic NCQM}

In the usual formulations of NCQM, interpreted here as relativistic theories, the coordinates~${\mathbf x}^\mu$ and their conjugate momenta ${\mathbf p}_\mu$ are operators acting in a Hilbert space ${\cal H}$  satisfying the
fundamental commutation relations given in Section~\ref{section2}, we can def\/ine the operator
\begin{gather*}
%\label{i10}
{\mathbf G}_1={1\over2}\omega_{\mu\nu}{\mathbf L}^{\mu\nu}.
\end{gather*}
Note that, analogously to (\ref{32000}), it is possible to dynamically generate inf\/initesimal transformations on any operator ${\mathbf A}$, following the usual rule
$\delta {\mathbf A}=i[ {\mathbf A}, {\mathbf G}_1]$. For ${\mathbf X}^\mu$,
${\mathbf p}_\mu$ and ${\mathbf L}^{\mu\nu}$, given in~\eqref{16000} and~\eqref{29}, with spacetime coordinates, we have the following results
\begin{gather*}
\delta {\mathbf X}^\mu=\omega ^\mu_{\,\,\,\,\nu}{\mathbf X}^\nu,\qquad
\delta{\mathbf p}_\mu=\omega _\mu^{\,\,\,\,\nu}{\mathbf p}_\nu,\qquad
\delta {\mathbf L}^{\mu\nu}=\omega ^\mu_{\,\,\,\,\rho}{\mathbf L}^{\rho\nu}+ \omega ^\nu_{\,\,\,\,\rho}{\mathbf L}^{\mu\rho}.%\label{i11}
\end{gather*}
However, the physical coordinates fail to transform in the appropriate way. As can be seen, the same rule applied on ${\mathbf x}^\mu$ gives the result
\begin{gather}
\label{i12}
\delta {\mathbf x}^\mu=\omega ^\mu_{\,\,\,\,\nu}
\left({\mathbf x}^\nu+{1\over2}\theta^{\rho\nu}{\mathbf p}_\nu\right)-
{1\over2}\theta^{\mu\nu}\omega_{\nu\rho}{\mathbf p}^{\rho},
\end{gather}
which is a consequence of $\theta^{\mu\nu}$ not being transformed. Relation (\ref{i12}) probably will break  Lorentz symmetry in any reasonable  theory. The cure for  these problems can be obtained by conside\-ring~$\theta^{\mu\nu}$ as an operator in ${\cal H}$, and introducing its canonical momentum $\pi_{\mu\nu}$ as well. The price to be paid is that $\theta^{\mu\nu}$ will have to be associated with extra dimensions, as happens with the formulations appearing in~\cite{Carlson, Banerjee2,Morita,Haghighat,Carone,Ettefaghi,Saxell}.

Moreover, we have that the commutation relation
\begin{gather}
\label{i15}
[{\mathbf x}^\mu,\pi_{\rho\sigma}]=-{i\over2}\delta^{\mu\nu}_{\,\,\,\rho\sigma}{\mathbf p}_\nu
\end{gather}
is necessary  for algebraic consistency  under Jacobi identities. The set~(\ref{i15}) completes the algebra displayed in Section~\ref{section2}, namely, the DFRA algebra.  With this algebra in mind, we can generalize the expression for the total angular momentum, equations~(\ref{30}) and~(\ref{31}).

The framework constructed above permits consistently to write \cite{Gracia}
\begin{gather}
\label{i16}
{ \mathbf M}^{\mu\nu}= { \mathbf X}^\mu{\mathbf p}^\nu-{\mathbf X}^\nu{\mathbf p}^\mu-\theta^{\mu\sigma}\pi_\sigma^{\,\,\nu}+\theta^{\nu\sigma}\pi_\sigma^{\,\,\mu}
\end{gather}
and consider this object as the generator of the Lorentz group, since it not only closes in the appropriate algebra
\begin{gather}
\label{i17a}
[{\mathbf M}^{\mu\nu},{\mathbf M}^{\rho\sigma}]=i\eta^{\mu\sigma}{\mathbf M}^{\rho\nu}-i\eta^{\nu\sigma}{\mathbf M}^{\rho\mu}-i\eta^{\mu\rho}{\mathbf M}^{\sigma\nu}+i\eta^{\nu\rho}{\mathbf M}^{\sigma\mu} ,
\end{gather}
but it generates the  expected Lorentz transformations on the Hilbert space operators. Actually, for
$\delta {\mathbf A}=i[ {\mathbf A}, {\mathbf G}_2]$, with $ {\mathbf G}_2={1\over2}\omega_{\mu\nu}{\mathbf M}^{\mu\nu}$, we have that,
\begin{gather}
\delta {\mathbf x}^\mu=\omega ^\mu_{\,\,\,\,\nu}{\mathbf x}^\nu,\qquad
\delta {\mathbf X}^\mu=\omega ^\mu_{\,\,\,\,\nu}{\mathbf X}^\nu,\qquad
\delta{\mathbf p}_\mu=\omega _\mu^{\,\,\,\,\nu}{\mathbf p}_\nu,\qquad
\delta\theta^{\mu\nu}=\omega ^\mu_{\,\,\,\,\rho}\theta^{\rho\nu}+ \omega ^\nu_{\,\,\,\,\rho}\theta^{\mu\rho},\nonumber\\
\delta\pi_{\mu\nu}=\omega _\mu^{\,\,\,\,\rho}\pi_{\rho\nu}+ \omega _\nu^{\,\,\,\,\rho}\pi_{\mu\rho},\qquad
\delta {\mathbf M}^{\mu\nu}=\omega ^\mu_{\,\,\,\,\rho}{\mathbf M}^{\rho\nu}+ \omega ^\nu_{\,\,\,\,\rho}{\mathbf M}^{\mu\rho} ,\label{i18}
\end{gather}
which in principle should guarantee the Lorentz invariance of a consistent  theory.
We observe that this construction is possible because of the introduction of the canonical pair
${\mathbf\theta}^{\mu\nu}$, $\pi_{\mu\nu}$ as independent  variables.
This pair allows the building of an object like ${\mathbf M}^{\mu\nu}$ in~(\ref{i16}), which generates the transformations given just above  dynamically~\cite{Iorio} and not merely by taking into account the algebraic index content of the variables.

From the symmetry structure given above, we realize that actually the Lorentz generator~(\ref{i16}) can be written as the sum of two commuting objects,
\[
{ \mathbf M}^{\mu\nu}= { \mathbf M}_1^{\mu\nu}+{ \mathbf M}_2^{\mu\nu} ,
\]
where
\[
{ \mathbf M}_1^{\mu\nu}= { \mathbf X}^\mu{\mathbf p}^\nu-{\mathbf X}^\nu{\mathbf p}^\mu
\qquad \mbox{and}\qquad
{ \mathbf M}^{\mu\nu}_2=-\theta^{\mu\sigma}\pi_\sigma^{\,\,\nu}+\theta^{\nu\sigma}\pi_\sigma^{\,\,\mu},
\]
as in the usual addition of angular momenta. Of course both operators have to satisfy the Lorentz algebra. It is possible to f\/ind convenient representations that reproduce (\ref{i18}). In the sector
${\cal H}_1$ of ${\cal H}={\cal H}_1\otimes{\cal H}_2$ associated with  (${\mathbf X},{\mathbf p}$), it can be used the usual $4\times4$  matrix representation $D_1(\Lambda)=(\Lambda^\mu_{\,\,\,\alpha})$, such that, for instance
\[
 {\mathbf X'}^\mu=\Lambda^\mu_{\,\,\,\nu}{\mathbf X}^\nu.
\]
  For the sector of ${\cal H}_2$ relative to ($\theta,\pi$), it is possible to use the $6\times6$ antisymmetric product representation
\[
D_2(\Lambda)=\big(\Lambda^{[\mu}_{\,\,\,\alpha}\Lambda^{\nu]}_{\,\,\,\beta}\big),
\]
such that, for instance,
\[
{\mathbf \theta'}^{\mu\nu}=\Lambda^{[\mu}_{\,\,\,\alpha}\Lambda^{\nu]}_{\,\,\,\beta}\theta^{\alpha\beta}.
\]
The complete representation is given by $D=D_1\oplus D_2$.  In the inf\/initesimal case, $\Lambda^\mu_{\,\,\,\nu}=\delta^\mu_\nu+\omega^\mu_{\,\,\,\nu}$, and (\ref{i18}) are reproduced.
There are four Casimir invariant operators in this context and they are given by
\[
{\mathbf C_j}_{1}={\mathbf M_j}^{\mu\nu}{ \mathbf M_j}_{\mu\nu}
\qquad \mbox{and}\qquad { \mathbf C_j}_{2}=\epsilon_{\mu\nu\rho\sigma}{ \mathbf M_j}^{\mu\nu}{ \mathbf M_j}^{\rho\sigma} ,
\]
where $j=1,2$. We note that although the target space has $10=4+6$ dimensions, the symmetry group has only 6 independent parameters
and not the 45 independent parameters of the  Lorentz group in $D=10$.
As we said before, this $D=10$ spacetime comprises the four spacetime coordinates and the six $\theta$ coordinates.  In Section~\ref{section6} the structure of this extended space will become clearer.

Analyzing the Lorentz symmetry in NCQM following the lines above, once we introduce an appropriate theory, for instance, given by a scalar action.  We know, however, that the  elementary particles are classif\/ied according to the eigenvalues of the Casimir operators of the inhomogeneous Lorentz group. Hence, let us extend this approach to the Poincar\'{e} group ${\cal P}$. By considering the operators presented here, we can in principle consider
\[
{\mathbf G}_3={1\over2}\omega_{\mu\nu}{\mathbf M}^{\mu\nu}-a^\mu{\mathbf p}_\mu+{1\over2}b_{\mu\nu}\pi^{\mu\nu}
\]
as the generator of some   group  ${\cal P}'$, which has the Poincar\'{e} group as a subgroup.  By following the same rule as the one used in the obtainment of (\ref{i18}), with  ${\mathbf G}_2$ replaced by ${\mathbf G}_3$, we arrive at the set of transformations
\begin{gather}
\delta {\mathbf X}^\mu=\omega ^\mu_{\,\,\,\,\nu}{\mathbf X}^\nu+a^\mu,\qquad
\delta{\mathbf p}_\mu=\omega _\mu^{\,\,\,\,\nu}{\mathbf p}_\nu,\qquad
\delta\theta^{\mu\nu}=\omega ^\mu_{\,\,\,\,\rho}\theta^{\rho\nu}+ \omega ^\nu_{\,\,\,\,\rho}\theta^{\mu\rho}+b^{\mu\nu},\nonumber\\
\delta\pi_{\mu\nu}=\omega _\mu^{\,\,\,\,\rho}\pi_{\rho\nu}+ \omega _\nu^{\,\,\,\,\rho}\pi_{\mu\rho},\qquad
\delta {\mathbf M}_1^{\mu\nu}=\omega ^\mu_{\,\,\,\,\rho}{\mathbf M}_1^{\rho\nu}+ \omega ^\nu_{\,\,\,\,\rho}{\mathbf M}_1^{\mu\rho}+a^\mu{\mathbf p}^\nu-a^\nu{\mathbf p}^\mu,\nonumber\\
\delta {\mathbf M}_2^{\mu\nu}=\omega ^\mu_{\,\,\,\,\rho}{\mathbf M}_2^{\rho\nu}+ \omega ^\nu_{\,\,\,\,\rho}{\mathbf M_2}^{\mu\rho}+b^{\mu\rho}\pi_\rho^{\,\,\,\,\nu}+ b^{\nu\rho}\pi_{\,\,\,\rho}^{\mu},\qquad
\delta {\mathbf x}^\mu=\omega ^\mu_{\,\,\,\,\nu}{\mathbf x}^\nu+a^\mu+{1\over2}b^{\mu\nu}{\mathbf p}_\nu.\label{i19a}
\end{gather}
We observe that there is an unexpected term in the last one of (\ref{i19a}) system. This is a consequence of the coordinate operator in (\ref{16000}), which is a nonlinear combination of operators that act on~${\cal H}_1$ and~${\cal H}_2$.

The action of ${\cal P}'$ over the Hilbert space operators is in some sense equal to the action of the
Poincar\'{e} group with an additional translation operation on the ($\theta^{\mu\nu}$) sector.
All its generators close in an algebra under commutation, so ${\cal P}'$ is a well def\/ined group of transformations. As a~matter of fact, the commutation of two  transformations closes in the algebra
\begin{gather}
\label{i19aa}
[\delta_2,\delta_1] {\mathbf y}=\delta_3 {\mathbf y},
\end{gather}
where ${\mathbf y}$ represents any one of the operators appearing in~(\ref{i19a}). The parameters composition rule is given by
\begin{gather}
\omega^\mu_{3\nu}=\omega^\mu_{1\alpha}\omega^\alpha_{2\nu}-\omega^\mu_{2\alpha}\omega^\alpha_{1\nu},\qquad
a_3^\mu=\omega^\mu_{1\nu}a_2^\nu-\omega^\mu_{2\nu}a_1^\nu,\nonumber\\
b_3^{\mu\nu}=\omega^\mu_{1\rho}b_2^{\rho\nu}-\omega^\mu_{2\rho}b_1^{\rho\nu}-\omega^\nu_{1\rho}b_2^{\rho\mu}+
\omega^\nu_{2\rho}b_1^{\rho\mu}. \label{i19b}
\end{gather}

If we consider the operators acting  only on ${\cal H}_1$, we verify that they transform standardly under the Poincar\'{e} group ${\cal P}$ in $D=4$, whose generators are ${\mathbf p}^\mu$ and ${\mathbf M}_1^{\mu\nu}$. As it is well known, it is formed by the semidirect product between the Lorentz group $L$  in $D=4$ and the translation group $T_4$, and have two Casimir invariant operators
 ${\mathbf C}_1={\mathbf p}^2$ and ${\mathbf C}_2={\mathbf s}^2$, where
${\mathbf s}_\mu={1\over2}\epsilon_{\mu\nu\rho\sigma}{\mathbf M}_1^{\nu\rho}{\mathbf p}^\sigma$  is the Pauli--Lubanski vector. If we include in ${\mathbf M}_1$ terms associated with spin, we will keep the usual classif\/ication of the elementary particles based on those invariants.
A representation for ${\cal P}$ can be given by the $5\times5$ matrix{\samepage
\begin{gather*}%\label{P}
D_3(\Lambda,A)= \left( \begin{array}{cc}
        \Lambda^\mu_{\,\,\,\nu}&A^\mu\\
	0&1 \end{array} \right)
\end{gather*}
acting in the 5-dimensional  vector ${{\mathbf X}^\mu \atopwithdelims ( ) 1}$.}

Considering the operators acting on ${\cal H}_2$, we f\/ind a similar structure.  Let us call the corresponding  symmetry group as $G$. It has as generators the operators $\pi^{\mu\nu}$ and ${\mathbf M}_2^{\mu\nu}$. As one can verify,  ${\mathbf C}_3=\pi^2$ and ${\mathbf C}_4={\mathbf M}_2^{\mu\nu}\pi_{\mu\nu}$ are the corresponding  Casimir operators. $G$  can be   seen as the semidirect product of the Lorentz group and the translation group $T_6$. A possible representation uses the antisymmetric $6\times6$  representation $D_2(\Lambda)$ already discussed, and is given by the $7\times7$ matrix
\begin{gather*}%\label{G}
D_4(\Lambda,B)= \left( \begin{array}{cc}
\Lambda^{[\mu}_{\,\,\,\alpha}\Lambda^{\nu]}_\beta&B^{\mu\nu}\\
	0&1 \end{array} \right)
\end{gather*}
acting in the 7-dimensional  vector ${\theta^{\mu\nu}\atopwithdelims ( )1}$.
Now we see that the complete group ${\cal P}'$ is just the product of ${\cal P}$ and $G$. It has a $11\times11$ dimensional representation given by
\begin{gather}\label{aP'}
D_5(\Lambda,A, B)= \left( \begin{array}{ccc}
                \Lambda^\mu_{\,\,\,\nu} & 0 & A^\mu \\
                0 & \Lambda^{[\mu}_{\,\,\,\alpha}\Lambda^{\nu]}_\beta & B^{\mu\nu}\\
	              0 & 0 & 1 \end{array} \right)
\end{gather}
acting in the 11-dimensional column vector
\[
\left( \begin{array}{c} X^\mu\\ \theta^{\mu\nu}\\ 1 \end{array} \right).
\]
A group element needs $6+4+6$ parameters to be determined and ${\cal P}'$ is a subgroup of the full Poincar\'{e}   group $P_{10}$ in $D=10$.  Observe that an element of ${\cal P}_{10}$ needs 55 parameters to be specif\/ied. Here, in the inf\/initesimal case, when $A$ goes to $a$, $B$ goes to $b$ and $\Lambda^\mu_{\,\,\,\nu}$ goes to $\delta^\mu_\nu+\omega^\mu_{\,\,\,\nu}$, the transformations (\ref{i19a}) are obtained from the action of (\ref{aP'}) def\/ined above.
It is clear that ${\mathbf C}_1$, ${\mathbf C}_2$, ${\mathbf C}_3$ and ${\mathbf C}_4$ are the Casimir operators of
${\cal P}'$.

So far we have been considering a possible algebraic structure among operators in ${\cal H}$ and  possible sets of transformations for these operators. The choice of an specif\/ic theory, however, will give  the mandatory
criterion for selecting among these sets of transformations, the one that gives the dynamical symmetries of the action. If the considered theory is not invariant under the~$\theta$ translations, but it is by Lorentz transformations and $x$ translations, the set of the symmetry transformations on the generalized coordinates will be given by \eqref{i19a} but ef\/fectively conside\-ring~$b_{\mu\nu}$ as vanishing, which implies that~$P'$, with this condition, is  dynamically contracted to the Poincar\'{e} group.  Observe, however, that   $\pi_{\mu\nu}$ will be yet a relevant operator, since~${\mathbf M}_{\mu\nu}$ depends on it in the representation here adopted.
An important point related with the dynamical action of ${\cal P}$ is that it conserves the quantum conditions~(\ref{04000}).

Now we will consider some  points concerning some actions which furnish  models for a~re\-la\-tivistic NCQM  in order to derive their equations of motion and to display their symmetry content.

\subsection{Actions}

As discussed in the previous section, in NCQM the physical coordinates do not commute and their eigenvectors can not be used in order to form a basis in ${\cal H}={\cal H}_1+{\cal H}_2$.
This does not occur with the shifted coordinate operator ${\mathbf X}^\mu$ due to \eqref{10}, \eqref{11} and (\ref{i6}). Consequently their eigenvectors can be used in the construction of such a basis. Generalizing what has been done in \cite{Amorim1},  it is possible to introduce a
 coordinate basis $|X',\theta'\rangle=|X'\rangle \otimes |\theta'\rangle$ in such a way that
\begin{gather}\label{b1}
 {\mathbf X}^\mu|X',\theta'\rangle = {X'}^\mu|X',\theta'\rangle, \qquad
{\mathbf\theta}^{\mu\nu} |X',\theta'\rangle = {\theta'}^{\mu\nu}|X',\theta'\rangle
\end{gather}
satisfying usual orthonormality and completeness relations. In this basis
\begin{gather}
\label{b2}
\langle  { X}',{ \theta}'|{\mathbf p}_\mu|{ X}'',{ \theta}''\rangle= -i{\frac{\partial}{\partial X'^\mu}}\delta^{4} (X'-X'')\delta^{6}({\theta}'-{\theta}'')
\end{gather}
and
\begin{gather}
\label{b3}
\langle { X}',{ \theta}'|\pi_{\mu\nu}|{ X}'',{ \theta}''\rangle= -i\delta^{4} (X'-X''){\frac{\partial}{\partial \theta'^{\mu\nu}}}\delta^{6}({\theta}'-{\theta}'')
\end{gather}
implying that both momenta acquire a derivative realization.

A physical state $|\phi\rangle$,  in the coordinate basis def\/ined above, will be represented by the wave function $\phi(X',\theta')=\langle X',\theta'|\phi\rangle$
satisfying some wave equation that we assume that can be derived from an action, through a variational principle.
As it is well known,  a  direct route for constructing  an ordinary relativistic free quantum theory is to impose that the physical states are annihilated by the mass shell condition
\begin{gather}
\label{b4}
({\mathbf p}^2+m^2)|\phi\rangle =0
\end{gather}
constructed with the Casimir operator ${\mathbf C}_1={\mathbf p}^2$. In the coordinate representation, this gives the Klein--Gordon equation.
The same result is obtained from the quantization of the classical relativistic particle, whose action is invariant under reparametrization~\cite{Dirac}. There the generator of
the reparametrization symmetry is the constraint $({\mathbf p}^2+m^2)\approx0$. Condition~(\ref{b4}) is then interpreted as the one that selects the physical states, that must be invariant under gauge (reparametrization) transformations.
In the NC case, besides (\ref{b4}), it is reasonable to assume as  well that the second condition
\begin{gather}
\label{b5}
(\pi^2+\Delta)|\phi\rangle =0
\end{gather}
must be imposed on the physical states, since it is also an invariant, and it is not af\/fected by the evolution generated by (\ref{b4}). It can be shown that in the underlying classical theory \cite{NEXT}, this condition is also associated with a f\/irst-class constraint, which generates gauge transformations, and so (\ref{b5}) can also be seen as selecting gauge invariant states.
In (\ref{b5}), $\Delta$ is some constant with dimension of $M^4$, whose sign and value depend if $\pi$ is space-like, time-like or null. Both equations
permit to construct a generalized plane wave solution
\[
\phi(X',\theta')\equiv\langle X',\theta'|\phi\rangle \sim \exp\left(ik_\mu{X'}^\mu+{i\over2}K_{\mu\nu}{\theta'}^{\mu\nu}\right),
\]
where $k^2+m^2=0$ and $K^2+\Delta=0$.
In coordinate representation given by equations (\ref{b1})--(\ref{b3}), the equation~(\ref{b4}) gives just the Klein--Gordon equation
\begin{gather*}
%\label{b6}
\big(\Box_X-m^2\big)\phi(X',\theta')=0
\end{gather*}
while (\ref{b5}) gives the supplementary equation
\begin{gather}
\label{b7}
(\Box_{\theta}-\Delta)\phi(X',\theta')=0,
\end{gather}
where
\begin{gather} \label{bb1}
\Box_X= \partial^\mu\partial_\mu ,
\end{gather}
with
\begin{gather} \label{bb2}
\partial_\mu={{\partial}\over{\partial {X'}^\mu}} .
\end{gather}
 Also
\begin{gather} \label{bb3}
\Box_{\theta}={1\over2}\partial^{\mu\nu}\partial_{\mu\nu} ,
\end{gather}
with
\begin{gather} \label{bb4}
\partial_{\mu\nu}={{\partial}\over{\partial {\theta'}^{\mu\nu}}} .
\end{gather}
Both equations can be derived from the action
\begin{gather}
\label{bb8}
S = \int d^{4} X' d^{6}\theta' \,\Omega(\theta') \left\{ {1\over2}( \partial^\mu\phi\partial_\mu\phi+m^2 \phi^2) -\Lambda( \Box_\theta -\Delta)\phi\right\}.
\end{gather}
In (\ref{bb8}) $\Lambda$ is a Lagrange multiplier necessary to impose
condition (\ref{b7}). $\Omega(\theta')$ can be seen as a simple constant $\theta_0^{-6}$ to keep the usual dimensions of the  f\/ields as $S$ must be dimensionless in natural units, as an even weight function as the one appearing
in~\cite{Carlson, Banerjee2,Morita,Haghighat,Carone,Ettefaghi,Saxell} used to make the connection between the formalism in $D=4+6$ and the usual one in $D=4$ after the integration in $\theta'$, or a distribution used to impose further conditions as those appearing in~(\ref{04000}) and adopted in~\cite{DFR}.

A model not involving Lagrangian multipliers, but two of the Casimir operators of ${\cal P}'$, ${\mathbf C}_1={\mathbf p}^2$ and ${\mathbf C}_3=\pi^2$, is given by
\begin{gather}
\label{b8a}
S = \int d^{4} X' d^{6}\theta' \Omega(\theta')  {1\over2}\left\{ \partial^\mu\phi\partial_\mu\phi + {{\lambda^2}\over4} \partial^{\mu\nu}\phi\partial_{\mu\nu}\phi   +m^2 \phi^2\right\} ,
\end{gather}
where $\lambda$ is a parameter with dimension of length, as the Planck length, which has to be introduced by dimensional reasons. If it goes to zero one essentially obtains the Klein--Gordon action. Although, the Lorentz-invariant weight function $\Omega(\theta)$ does not exist we can keep it temporally in some integrals in order to guarantee explicitly their existence.  By borrowing the Dirac matri\-ces~$\Gamma_A$, $A=0,1,\dots,9$, written for spacetime $D=10$, and identifying the tensor indices with the six last values of~$A$, it is also possible to construct the ``square root'' of the equation of motion derived from~(\ref{bb8}), obtaining a generalized Dirac theory involving spin and noncommutativity~\cite{NEXT}.  It is an object of current investigation.

Next we will construct the equations of motion and analyze the Noether's theorem derived for general theories def\/ined in $x+\theta$ space, and specif\/ically  for the action (\ref{bb8}), considering $\Omega(\theta)$ as a well behaved function.

\subsection{Equations of motion and Noether's theorem}

Let us consider the action
\begin{gather}
\label{c1}
S=\int_R d^{4} x d^{6}\theta \,\Omega(\theta) {\cal L}(\phi^i,\partial_\mu\phi^i,\partial_{\mu\nu}\phi^i,x, \theta) ,
\end{gather}
relying on a set of f\/ields $\phi^i$, the derivatives with respect to ${x}^\mu$ and ${\theta}^{\mu\nu}$ and the
coordinates ${x}^\mu$ and ${\theta}^{\mu\nu}$ themselves. From now on we will use $x$ in place of $X'$ and $\theta$ in place of $\theta'$ in order to simplify the notation. Naturally the f\/ields~$\phi^i$ can be functions of ${x}^\mu$ and ${\theta}^{\mu\nu}$. The index~$i$ permits to treat~$\phi$ in a general way. In~(\ref{c1}) we consider, as in~(\ref{bb8}), the integration element modif\/ied by the introduction of~$\Omega(\theta)$.

By assuming that $S$ is stationary for an arbitrary variation  $\delta\phi^i$  vanishing on the boundary $\partial R$ of the region of integration $R$, we can write the Euler--Lagrange equation as
\begin{gather}
\label{c2}
\Omega\left({{\partial{\cal L}}\over{\partial\phi^i}}-\partial_\mu{{\partial{\cal L}}\over{\partial\partial_\mu\phi^i}}\right)-\partial_{\mu\nu}\left(\Omega{{\partial{\cal L}}\over{\partial\partial_{\mu\nu}\phi^i}}\right)=0 .
\end{gather}
We will treat the variations  $\delta x^\mu$, $\delta\theta^{\mu\nu}$ of the generalized coordinates and $\delta\phi^i$ of the f\/ields such that the integrand transforms as a total divergence in the $x+\theta$ space,   $\delta(\Omega\,{\cal L})=\partial_\mu(\Omega\,S^\mu)+\partial_{\mu\nu}(\Omega\,S^{\mu\nu})$. Then the Noether's theorem assures that, on shell, or when~(\ref{c2}) is satisf\/ied, there is a  conserved current $(j^\mu,j^{\mu\nu})$ def\/ined by
\begin{gather}
j^\mu = {{\partial{\cal L}}\over{\partial\partial_\mu\phi^i}}\delta\phi^i+{\cal L}\delta x^\mu,\qquad
j^{\mu\nu} = {{\partial{\cal L}}\over{\partial\partial_{\mu\nu}\phi^i}}\delta\phi^i+{\cal L}\delta\theta^{\mu\nu},\label{c3}
\end{gather}
such that
\begin{gather}
\label{c4}
\Xi= \partial_\mu(\Omega  j^\mu)+\partial_{\mu\nu}(\Omega j^{\mu\nu})
\end{gather}
vanishes.
The corresponding charge
\begin{gather}
\label{c5}
Q=  \int d^{3} x d^{6}\theta\,\Omega(\theta) j^0
\end{gather}
is independent of the ``time'' ${x}^0$. On the contrary, if there exists a conserved current like~(\ref{c3}), the action (\ref{c1}) is invariant under the corresponding symmetry transformations. This is just a trivial extension of the usual version of Noether's theorem~\cite{Iorio} in order to include $\theta^{\mu\nu}$ as independent coordinates, as well as a modif\/ied integration element due to the presence of $\Omega(\theta)$.  Notice that $\Omega$ has not been included in current def\/inition~(\ref{c3}) because it is seen as part of the element of integration,  but it is present in~(\ref{c4}), which is the relevant divergence. It is also inside the charge~(\ref{c5}) since the charge is an integrated quantity.

Let us use the equations from (\ref{c1}) until (\ref{c4}) to the simple model given by (\ref{b8a}). The Lagrange equation reads
\begin{gather}
\label{c6000}
{{\delta S}\over{\delta\phi}} = - \Omega \big(\Box - m^2\big)\phi-{{\lambda^2}\over2}\partial_{\mu\nu}(\Omega \partial^{\mu\nu}\phi) = 0
\end{gather}
and (\ref{c4}) can be written as
\begin{gather}
\Xi = \partial_\mu\left\{\Omega \partial^\mu\phi \delta\phi+{{\Omega}\over{2}}
\left(\partial_\alpha\phi\partial^\alpha\phi+{{\lambda^2}\over4}
\partial_{\alpha\beta}\phi\partial^{\alpha\beta}\phi+m^2\phi^2\right)\delta x^\mu\right\}\nonumber\\
\phantom{\Xi =}{} +\partial_{\mu\nu}\left\{\Omega \lambda^2\partial^{\mu\nu}\phi \delta\phi
+{{\Omega}\over{2}}\left(\partial_\alpha\phi\partial^\alpha\phi+{{\lambda^2}\over4}
\partial_{\alpha\beta}\phi\partial^{\alpha\beta}\phi+m^2\phi^2\right)\delta\theta^{\mu\nu}\right\} . \label{c61}
\end{gather}

Before using (\ref{c61}) we observe that the transformation
\begin{gather}
\label{i19c}
\delta \phi=-(a^\mu+\omega^\mu_{\,\,\,\nu}x^\nu)\,\partial_\mu\phi-{1\over2}(b^{\mu\nu}
+2\omega^\mu_{\,\,\,\rho}\theta^{\rho\nu})\,\partial_{\mu\nu}\phi
\end{gather}
 closes in an algebra, as in \eqref{i19aa}, with the same composition rule def\/ined in (\ref{i19b}). The above equation def\/ines  how a scalar f\/ield transforms in the $x+\theta$ space under the action of ${\cal P}'$.

Let us now study a rigid $x$-translation, given by
\begin{gather}
\delta_a x^\mu = a^\mu,\qquad
  \delta_a\theta^{\mu\nu} = 0,\qquad
\delta_a\phi = -a^\mu\partial_\mu\phi,\label{c62}
\end{gather}
where $a^\mu$ are constants.
We see from (\ref{c61}) and (\ref{c62}) that
\begin{gather*}
%\label{c63}
\Xi_a=a^\mu\partial_\mu\phi {{\delta S}\over{\delta\phi}}
\end{gather*}
vanishing on shell, when (\ref{c6000}) is valid.

For a rigid $\theta$-translation, we have that,
\begin{gather*}
\delta_b x^\mu= 0,\qquad
  \delta_b\theta^{\mu\nu} = b^{\mu\nu},\qquad
\delta_b\phi = -{1\over2}b^{\mu\nu}\partial_{\mu\nu}\phi,%\label{c64}
\end{gather*}
where $b^{\mu\nu}$ are constants, we can write
\begin{gather*}
%\label{c65}
\Xi_b={1\over2}b^{\mu\nu}\left(\partial_{\mu\nu}\phi {{\delta S}\over{\delta\phi}}+{\cal L}\partial_{\mu\nu}\Omega\right) .
\end{gather*}

The f\/irst term on the right vanishes on shell but the second one depends on the form of $\Omega$. Later we will comment this point.
To end, let us consider a Lorentz transformation, given by
\begin{gather*}
\delta_\omega x^\mu = \omega^{\mu}_{\,\,\,\nu} x^\nu, \qquad
\delta_\omega\theta^{\mu\nu} = \omega^{\mu}_{\,\,\,\rho}\theta^{\rho\nu}+\omega^{\nu}_\rho\theta^{\mu\rho},\nonumber\\
\delta_\omega\phi = -\big(\omega^\mu_{\,\,\,\nu} x^\nu\partial_\mu+ \omega^{\mu}_{\,\,\,\rho}\theta^{\rho\nu}\partial_{\mu\nu}\big) \phi%\label{c66}
\end{gather*}
with constant and antisymmetric $\omega^\mu_{\,\,\,\nu}$. We obtain
\begin{gather*}
%\label{c67}
\Xi_\omega=\omega^{\mu}_{\,\,\,\nu}\,{{\delta S}\over{\delta\phi}}(x^\nu\partial_\mu\phi+\theta^{\nu\rho}\partial_{\mu\rho}\phi)+{\cal L}\partial_{\mu\nu}\Omega\omega^\mu_{\,\,\,\alpha}\theta^{\alpha\nu}.
\end{gather*}
The f\/irst term in the above expression vanishes on shell and the second one also vanishes if $\Omega$ is a scalar under Lorentz transformations and depends only on $\theta$.

In a complete theory where other contributions for the total action would be present, the symmetry under $\theta$ translations could be broken by dif\/ferent reasons, as  in what follows, in the case of the NC $U(1)$ gauge theory. In this situation ${\cal P}$ could be the symmetry group of the complete theory even considering $\Omega(\theta)$ as a constant.

\subsection{Considerations about the twisted Poincar\'e symmetry and DFRA space}

The twisted Poincar\'e (TP) algebra describes the symmetry of NC spacetime whose coordinates obey the commutation relation of a canonical type like~\eqref{a1}.  It was suggested \cite{tese} as a substitute for the Poincar\'e symmetry in f\/ield theories on the NC spacetime.  In this formalism point of view, the Moyal product~\eqref{a11.22} is obtained as a twisted product of a module algebra of the TP algebra.  This result shows the TP invariance of the NC f\/ield theories.

This Poincar\'e algebra ${\cal P}$ for commutative QFT is given by the so-called Lie algebra composed by the ten generators of the Poincar\'e group \cite{tese}
\begin{gather}
\left[ {\mathbf P}_\mu , {\mathbf P}_\nu \right]  =  0, \qquad
\left[ {\mathbf M}_{\mu\nu},{\mathbf P}_{\rho} \right]  =  - i  (g_{\mu\rho} {\mathbf P}_\nu - g_{\nu\sigma} {\mathbf P}_\mu  ),  \nonumber \\
\left[ {\mathbf M}_{\mu\nu},{\mathbf M}_{\rho\sigma} \right]  =  - i ( g_{\mu\rho} {\mathbf M}_{\nu\sigma}\ - g_{\mu\sigma} {\mathbf M}_{\nu\rho} - g_{\nu\rho} {\mathbf M}_{\mu\sigma} - g_{\nu\sigma} {\mathbf M}_{\mu\rho} ) , \label{A}
\end{gather}
where the matrix ${\mathbf M}$ is antisymmetric ${\mathbf M}_{\mu\nu} = - {\mathbf M}_{\nu\mu}$.  The generators ${\mathbf M}_{\mu\nu}$ form a closed subalgebra, which is the Lie algebra of the Lorentz group.  The generators of the Lorentz group can be divided into the group for boosts ${\mathbf K}_i = {\mathbf M}_{0i}$, the spatial rotations group ${\mathbf J}_i = \frac12 \epsilon_{ijk} {\mathbf M}_{jk}$ and the generators of translations ${\mathbf P}_\mu$, the last ones form a commutative subalgebra of the Poincar\'e algebra (the translation subgroup is Abelian).  The presence of the imaginary unit $i$ indicates that the generators of the Poincar\'e group are Hermitian.

The Poincar\'e algebra generators are represented by
\begin{gather*}
P_\mu  =  \int   d^d x\, T_{0\mu}(x) , \qquad
M_{\mu\nu}  =  \int   d^d x\, [ x_\mu T_{0\nu}(x) - x_\nu T_{0\mu} (x) ]  ,
\end{gather*}
where $T_{0\mu} (x) = \frac12 [ \pi(x) \partial_\mu \phi(x) + \partial_\mu \phi(x) \pi(x) ] - g_{0\mu} {\cal L}(x)$ and $\pi=\partial_0\,\phi$ is the canonical momentum of~$\phi^0$.  It can be demonstrated that with this structure we can construct the representation of the Poincar\'e algebra in terms of the Hopf algebraic structure~\cite{abe,twisted}.

\subsubsection{The twisted Poincar\'e algebra}

During some time, the problem of the Lorentz symmetry breaking was ignored and the investigations in NCQFT were performed by dealing with the full representation content of the Poincar\'e algebra.

Chaichian et al.\ in \cite{wess2} analyzed a solution to the problem utilizing the form of a twisted Poincar\'e symmetry.  The introduction of a twist deformation of the universal enveloping algebra of the Poincar\'e algebra provided a new symmetry.  The representation content of this twisted algebra is the same as the representation content of the usual Poincar\'e algebra~\cite{tese}.

To obtain the TP algebra we use the standard  Poincar\'e algebra and its representation space and twist them.  For example, in this twisted algebra the energy-momentum tensor is $T_{0\mu}$ is given by
\begin{gather*}
T_{0\mu} =  \sum^\infty_{n=0}\left(-\frac12 \right)^n \frac{1}{n!} \theta^{i_1 j_1}\cdots\theta{i_n j_n} \partial_{i_1}\cdots \partial_{i_n}  \nonumber \\
\phantom{T_{0\mu} =}{}\times \left[ \frac12  ( \pi \star \partial_\mu \phi(x) + \partial_\mu \phi(x) \star \pi(x) ) - g_{0\mu} {\cal L}\right] P_{j_1}\cdots P_{j_n} ,
\end{gather*}
where $P_{j_n}$ are the generators of translation in NCQFT.

It is important to notice that $\pi$, $\phi$, ${\cal L}$ and $P$ are elements embedded in a NC space.  The resulting operators satisfy commutation relations of Poincar\'e algebra~(\ref{A}).  However, it can be shown~\cite{abe,twisted} that some identities involving momenta and the Hamiltonian in commutative and NCQFT are preserved in a deformed NCQFT.  Also, we can use the same Hilbert space to represent the f\/ield operator for both commutative and deformed NCQFT~\cite{abe,twisted}.  The action of generators of a Lorentz on a NC f\/ield transformation is dif\/ferent from the commutative one.  And this action has an exponential form in order to give a f\/inite Lorentz transformation.  This is, in a nutshell, the structure of the Poincar\'e symmetry or of the Hopf algebra represented on a deformed NCQFT.  The next step is to twist this algebra and its representation space.  The Poincar\'e algebra~${\cal P(A)}$ has a subalgebra (commutative) composed by the translation generators~$P_\mu$
\begin{gather*} %\label{B}
{\cal F} =\exp\left( \frac{i}{2} \theta^{\mu\nu} P_\mu P_\nu\right),
\end{gather*}
where $\theta^{\mu\nu}$ is a real constant antisymmetric matrix.  This twist operator obviously satisf\/ies the twist conditions that preserve the Hopf algebra structure~\cite{tese}.

Explicitly let us write the twist operator such as
\begin{gather}
{\cal F} =\exp\left[-\frac{i}{2} \theta^{\mu\nu} \frac{\partial}{\partial x^\mu} \otimes \frac{\partial}{\partial x^\nu}\right],\qquad
{\cal F}^{-1} =\exp\left[\frac{i}{2} \theta^{\mu\nu} \frac{\partial}{\partial x^\mu} \otimes
\frac{\partial}{\partial x^\nu} \right] ,\label{C}
\end{gather}
where here $\frac{\partial}{\partial x^\mu}$ and $\frac{\partial}{\partial x^\nu}$ are generally def\/ined vector f\/ields on space or spacetime.  The twisting result is the twisted Poincar\'e algebra.  It can be shown~\cite{abe,twisted} that the Poincar\'e covariance of a commutative QFT implies the twisted Poincar\'e covariance of a deformed NCQFT.  In this deformed NCQFT the symmetry is described by a quantum group.

An unavoidable comparison with the Moyal $\star$-product can be made and the conclusion is that they are in fact the same NC product of functions.  Hence, the NCQFT constructed with Weyl quantization and Moyal $\star$-product possess the twisted Poincar\'e symmetry.  We can say that, in NC theories, relativistic invariance means invariance under the twisted Poincar\'e transformations.

\subsubsection{The DFRA analysis}

As we said through the sections above, the DFRA structure constructs an extension of the Poincar\'e group ${\cal P}'$, which has the Poincar\'e group ${\cal P}$ as a subgroup.  Then, obviously, the theories considered have to be invariant under both ${\cal P}$ and ${\cal P}'$.

As the considerations described in this section the Poincar\'e algebra ${\cal P}$ has an Abelian subalgebra which allows us to construct a twist operator ${\cal F}$ depicted in (\ref{C}) which is an element of the quantum group theory.

We believe that the subalgebra of the extended Poincar\'e algebra ${\cal P}'$ also permits the elaboration of a kind of extended twist operator ${\cal F}'$.

This extended twist operator in the DFRA framework has to be able to reproduce the new deformed generators.  Having the structure displayed in~(\ref{C}) in mind we have to formulate this new twist operator adding the $(\theta,\pi)$ sector terms in order to have the form of~(\ref{C}) with the twist operator as a special case.  The formulation of this DFRA twist operator is beyond the scope of this review paper and is a target for further investigations.

\section{Fermions and  noncommutative theories}\label{section5}

By using the DFRA framework where the object of noncommutativity $\theta^{\mu\nu}$ represents independent degrees of freedom, we will explain here the symmetry properties of
an extended $x+\theta$ spacetime, given by the group ${\cal P}'$, which has the Poincar\'{e} group ${\cal P}$ as a subgroup.  In this section we use the DFRA algebra to introduce a generalized Dirac equation, where the fermionic f\/ield depends not only  on the ordinary coordinates but  on $\theta^{\mu\nu}$ as well.
The dynamical symmetry content of such fermionic theory will be discussed now and it is shown that its action is invariant under  $\cal P'$.

In the last sections above, we saw that the DFRA algebra implemented, in a  NCQM framework\footnote{In~\cite{Amorim1} and~\cite{Amorim4} it is possible to f\/ind a large amount of references concerning NC quantum mechanics.},
%  \cite{NCQM},
the Poincar\'{e} invariance as a dynamical symmetry~\cite{Iorio}. Of course this represents one among several possibilities of incorporating noncommutativity in quantum theories, we saw also that not only
the coordinates ${\mathbf x}^\mu$ and their conjugate momenta ${\mathbf p}_\mu$ are operators acting in a~Hilbert space ${\cal H}$, but also
$\theta^{\mu\nu}$ and their canonical momenta $\pi_{\mu\nu}$ are considered as Hilbert space operators as well. The proposed DFRA algebra is given by (\ref{a1}), \eqref{6a}--\eqref{11} and \eqref{14}.
Where all these relations above are consistent with all possible Jacobi identities by construction.

As said before, an important point is that, due to (\ref{a1}) the operator ${\mathbf x}^\mu$ can not be used to label possible basis in~${\cal H}$. However, as the components of ${\mathbf X}^\mu$ commute, we know from QM that their eigenvalues  can be used for such purpose. To simplify the notation, let us denote by $x$ and $\theta$ the eigenvalues of~${\mathbf X}$ and~${\mathbf\theta}$ in what follows. In~\cite{Amorim4} R.~Amorim considered these points with some detail and have proposed a way for constructing some actions representing possible f\/ield theories in this extended $x+\theta$ spacetime.  One of such actions has been given by
\begin{gather}
\label{b8bb}
S = - \int d^{4} x d^{6}\theta\,\Omega(\theta)  {1\over2}\left\{ \partial^\mu\phi\partial_\mu\phi + {{\lambda^2}\over4} \partial^{\mu\nu}\phi\partial_{\mu\nu}\phi   +m^2 \phi^2\right\},
\end{gather}
where $\lambda$ is a parameter with dimension of length, as the Planck length, which has to be introduced by dimensional reasons and $\Omega(\theta)$
is a scalar weight function used in~\cite{Carlson,Banerjee2,Morita,Haghighat,Carone,Ettefaghi,Saxell} in order to make the connection between the $D=4+6$
and the $D=4$ formalisms, where we used the def\/initions inequations~(\ref{bb1})--(\ref{bb4}) and $\eta^{\mu\nu}={\rm diag}(-1,1,1,1)$.

The corresponding Lagrange equation reads, analogously as in (\ref{c6000}),
\begin{gather}
\label{c6cc}
{{\delta S}\over{\delta\phi}} = \Omega\big(\Box - m^2\big)\phi+{{\lambda^2}\over2}\partial_{\mu\nu}(\Omega \partial^{\mu\nu}\phi) =0
\end{gather}
 and the action (\ref{b8bb}) is invariant under the transformation (\ref{i19c}),
where $\Omega$ is considered as a~constant. If $\Omega$ is a scalar function of $\theta$, the above transformation is only a symmetry of (\ref{b8bb}) (when $b^{\mu\nu}$ vanishes) which dynamically
transforms~${\cal P}'$ to~${\cal P}$~\cite{Amorim4}. We observe that
\noindent (\ref{i19c}) closes in an algebra, as in~(\ref{i19aa}), with the same composition rule def\/ined in~(\ref{i19b}). That equation def\/ines  how a scalar f\/ield transforms in the $x+\theta$ space under the action of~${\cal P}'$.

In what follows we are going to show how to introduce fermions in this $x+\theta$ extended space. To reach this goal, let us f\/irst observe that
$\cal P$' is a subgroup of the Poincar\'{e} group ${\cal P}_{10}$ in $D=10$. Denoting the indices $A,B,\dots$ as spacetime indices in $D=10$, $A,B,\ldots=0,1,\dots,9$, a~vector~$Y^A$ would transform under~${\cal P}_{10}$ as
\[
\delta Y^A=\omega^A_{\,\,\,B}Y^B+ \Delta^A ,
\]
where the $45$ $\omega$'s and $10$ $\Delta$'s are inf\/initesimal parameters. If one identif\/ies the last six $A,B,\dots$ indices with the macro-indices $\mu\nu$, $\mu,\nu,\ldots =0,1,2,3$, considered as antisymmetric quantities, the transformation relations given above are rewritten as
\begin{gather*}
\delta Y^\mu = \omega^\mu_{\,\,\,\nu}Y^\nu+{1\over2}\omega^\mu_{\,\,\,\alpha\beta}Y^{\alpha\beta}+\Delta^\mu,\qquad
 \delta Y^{\mu\nu} = \omega^{\mu\nu}_{\,\,\,\,\,\,\,\alpha}Y^\alpha
 +{1\over2}\omega^{\mu\nu}_{\,\,\,\,\,\alpha\beta}Y^{\alpha\beta}+\Delta^{\mu\nu}.%\label{81}
\end{gather*}
With this notation, the (diagonal) $D=10$ Minkowski metric is rewritten as $\eta^{AB}=(\eta^{\mu\nu},\eta^{\alpha\beta,\gamma\delta})$ and the ordinary Clif\/ford algebra $\{\Gamma^A,\Gamma^B\}=-2\eta^{AB}$  as
\begin{gather}
\{\Gamma^\mu,\Gamma^{\alpha\beta}\} = 0,\qquad
\{\Gamma^\mu,\Gamma^\nu\} = -2\eta^{\mu\nu},\qquad
\{\Gamma^{\mu\nu},\Gamma^{\alpha\beta}\} = -2\eta^{\mu\nu,\alpha\beta}.\label{82}
\end{gather}

This is just a heavy way of writing  usual $D=10$ relations~\cite{Strings}. Now,
by identifying $Y^A$ with $(x^\mu,{1\over\lambda}\theta^{\alpha\beta})$, where $\lambda$ is some parameter with length dimension, we see from the structure given above  that the allowed transformations in~$\cal P$' are
those of  ${\cal P}_{10}$, submitted to the conditions
\begin{gather*}
%\label{83}
\omega^{\mu\nu}_{\,\,\,\alpha} = \omega_{\mu\nu}^{\,\,\,\alpha}=0, \qquad
\omega^{\mu\nu}_{\,\,\,\alpha\beta} = 4 \omega^{[\mu}_{\,\,\,\,\alpha}\delta^{\nu]}_{\,\,\,\beta},\qquad
\Delta^\mu   = a^\mu,\qquad
\Delta^{\alpha\beta} = {1\over\lambda}b^{\alpha\beta}
\end{gather*}
obviously keeping the identif\/ication between $\omega^{AB}$ and $\omega^{\mu\nu}$ when $A=\mu$ and $B=\nu$. Of course we have now only $6$ independent $\omega$'s and $10$ $a$'s and~$b$'s. With the relations given above it is possible to extract the ``square root'' of the generalized Klein--Gordon
equation (\ref{c6cc})
\begin{gather}
\label{84}
\big(\Box+\lambda^2\Box_\theta-m^2\big)\phi=0
\end{gather}
assuming here that $\Omega$ is a constant. We will see in the next section with details that this equation can be interpreted as a dispersion relation in this $D=4+6$ spacetime. Hence, this last equation furnish just the generalized Dirac equation
\begin{gather}
\label{85}
\left[ i\left(\Gamma^\mu\partial_\mu+{\lambda\over2}\Gamma^{\alpha\beta}\partial_{\alpha\beta}\right)-m\right]\psi=0 .
\end{gather}

Let us apply from the left on (\ref{85}) the operator
\[
\left[ i\left(\Gamma^\nu\partial_\mu+{\lambda\over2}\Gamma^{\alpha\beta}\partial_{\alpha\beta}\right)+m\right].
\]
After using (\ref{82}) we observe that $\psi$ satisf\/ies the generalized Klein--Gordon equation (\ref{84}) as well. The covariance of the generalized Dirac equation (\ref{85}) can also be proved. First we note that the operator
\begin{gather*}
%\label{86}
M^{\mu\nu}={i\over4}\big([\Gamma^\mu,\Gamma^\nu]+[\Gamma^{\mu\alpha},\Gamma^\nu_{\,\,\,\alpha}]\big)
\end{gather*}
 gives the desired representation for the $SO(1,3)$ generators, because it not only closes in the Lorentz algebra~\eqref{i17a}, but also satisf\/ies the commutation relations
\begin{gather*}
%\label{87}
 [\Gamma^\mu,M_{\alpha\beta}]=2i\delta^{\mu}_{[\alpha}\Gamma_{\beta]},\qquad
 [\Gamma^{\mu\nu},M_{\alpha\beta}]=2i\delta^{\mu}_{[\alpha}\Gamma^{\,\,\,\nu}_{\beta]}
 -2i\delta^{\nu}_{[\alpha}\Gamma^{\,\,\,\mu}_{\beta]}.
\end{gather*}

With these relations it is possible to prove that (\ref{85}) is indeed covariant under the Lorentz transformations given by
\begin{gather*}%\label{88}
\psi(x',\theta')=\exp\left(-{i\over2}\Lambda^{\mu\nu}M_{\mu\nu}\right)\psi(x,\theta).
\end{gather*}

By considering the complete $\cal P'$ group, we observe that the inf\/initesimal transformations of~$\psi$ are given by
\begin{gather}\label{89}
\delta\psi = - \left[(a^\mu+\omega^\mu_{\,\,\,\nu}x^\nu)\partial_\mu
+{1\over2}(b^{\mu\nu}+2\omega^\mu_{\,\,\,\rho}\theta^{\nu\rho})\partial_{\mu\nu}
+{i\over2}\omega^{\mu\nu}M_{\mu\nu}\right]\psi,
\end{gather}
which closes in the $\cal P'$ algebra with the same composition rule given by (\ref{i19b}), what can be shown after a little algebra. At last we can show that also here there are conserved Noether's currents associated with the transformation~(\ref{89}), once we observe that the equation (\ref{85})
can be derived from the action
\begin{gather}
\label{90}
S=\int d^{4} x d^{6}\theta\,\Omega(\theta)  \bar \psi \left[ i\left(\Gamma^\mu\partial_\mu+{\lambda\over2}\Gamma^{\alpha\beta}\partial_{\alpha\beta}\right)-m\right]\psi,
\end{gather}
where  we are considering that $\Omega=\theta_0^{-6}$ and $\bar \psi=\psi^\dag\Gamma^0$.
First  we note that (suppressing trivial $\theta_0^{-6}$ trivial factors)
\begin{gather*}
{{\delta^L S}\over{\delta\bar\psi}} = \left[ i\left(\Gamma^\mu\partial_\mu+{\lambda\over2}\Gamma^{\alpha\beta}\partial_{\alpha\beta}\right)-m\right]\psi,\qquad
{{\delta^R S}\over{\delta\psi}} = -\bar\psi\left[ i\left(\Gamma^\mu\overleftarrow\partial_\mu+{\lambda\over2}\Gamma^{\alpha\beta}\overleftarrow\partial_{\alpha\beta}\right)
+m\right],
%\label{91}
\end{gather*}
where $L$ and $R$ derivatives act from the left and right respectively. The current $(j^\mu,j^{\mu\nu})$, analogously as in (\ref{c3}), is here written as
\begin{gather}
\label{92}
 j^\mu={{\partial^R {\cal L}}\over{\partial\partial_\mu\psi}}\delta\psi+\delta\bar\psi {{\partial^L {\cal L}}\over{\partial\partial_\mu\bar\psi}}+
{\cal L}\delta x^\mu,\qquad
 j^{\mu\nu}={{\partial^R {\cal L}}\over{\partial\partial_{\mu\nu}\psi}}\delta\psi+
\delta\bar\psi {{\partial^L{\cal L}}\over{\partial\partial_{\mu\nu}\bar\psi}}+{\cal L}\delta \theta^{\mu\nu},
\end{gather}
 where
\begin{gather} \label{93}
\delta\bar\psi = -\bar\psi\left[\overleftarrow\partial_\mu(a^\mu+\omega^\mu_{\,\,\,\nu}x^\nu)
+\overleftarrow\partial_{\mu\nu}{1\over2}(b^{\mu\nu}
+2\omega^\mu_{\,\,\,\rho}\theta^{\nu\rho})-{i\over2}\omega^{\mu\nu}M_{\mu\nu}\right],
\end{gather}
 $\delta\psi$ is given in (\ref{89}) and $\delta x^\mu$ and $\delta\theta^{\mu\nu}$ have the same form found
in~\eqref{i19a}.   Using these last results one can show that,
\begin{gather} \label{94a}
\partial_\mu j^\mu+\partial_{\mu\nu}j^{\mu\nu}=-\left(\delta\bar\psi{{\delta^L S}\over{\delta\bar\psi}}+{{\delta^R S}\over{\delta\psi}}\delta\psi\right) ,
\end{gather}
which vanishes on shell, and hence the invariance of the action (\ref{90}) under ${\cal P}'$.
And we can conclude that it could be dynamically contracted to ${\cal P}$, preserving the usual Casimir invariant structure characteristic of ordinary quantum f\/ield theories.

Using (\ref{94a}) we can realize that there is a conserved charge
\begin{gather*}
%\label{95}
Q=\int d^3 x d^6 \theta\, j^0 ,
\end{gather*}
 for each one of the specif\/ic transformations encoded in~(\ref{92}).  Performing a simple $t$ derivative we have that
\[
\dot Q=-\int d^3 x d^6 \theta\, (\partial_i\,j^i+\partial_{\mu\nu}j^{\mu\nu}),
\]
vanishes as a consequence of the divergence theorem. By considering only $x^\mu$ translations, we can write $j^0=j^0_\mu a^\mu$, and consequently to def\/ine the
momentum operator
\[
P_\mu=-\int d^3x d^6\theta j^0_\mu.
 \]
 Analyzing $\theta^{\mu\nu}$ translations and Lorentz transformations, we can derive in a similar way an explicit form for the other generators of ${\cal P}'$, here denoted by~$\Pi_{\mu\nu}$ and~$J_{\mu\nu}$. Under an appropriate bracket structure, following the Noether's theorem, these conserved charges will generate the transformations~(\ref{89}) and~(\ref{93}).

Summarizing, we can say that we have been able to introduce fermions satisfying a generalized Dirac equation, which is covariant under the action of the extended Poincar\'{e} group ${\cal P}'$. This Dirac equation has been derived through a variational principle whose action is dynamically invariant under ${\cal P}'$. This can justify possible roles played by theories involving noncommutativity in a way compatible with Relativity. Of course this is just a little step toward a f\/ield theory quantization program in this extended $x+\theta$ spacetime, which is our next issue.

\section{Quantum complex scalar f\/ields and  noncommutativity}\label{section6}

So far we saw that in a f\/irst quantized formalism, $\theta^{\mu\nu}$ and its canonical momentum $\pi_{\mu\nu}$ are seen as operators  living in some Hilbert space.
This structure is compatible with DFRA algebra and it is invariant under an extended Poincar\'e group of symmetry.
In a second quantization scenario, we will reproduce in this section the results obtained in~\cite{aa}.
An explicit form for the extended Poincar\'e
generators will be presented and we will see the same algebra is generated via generalized Heisenberg relations.  We also introduce a source term and construct the general solution for the complex scalar f\/ields using the Green's function technique.

As we said before in a dif\/ferent way, at the beginning the original motivation of DFR to
study the relations (\ref{a1}), \eqref{triple} and \eqref{10} was the belief that an attempt of obtaining exact measurements involving spacetime localization could conf\/ine photons due to gravitational f\/ields.  This
phenomenon is directly related to~(\ref{a1}), \eqref{triple} and~\eqref{10} together with (\ref{04000}). In a somehow dif\/fe\-rent scenario, other relevant results are obtained in~\cite{Carlson,Haghighat,Carone,Ettefaghi,Morita,Saxell} relying on conditions~(\ref{a1}), \eqref{triple} and~\eqref{10}.  We saw that the value  of $\theta$ is used as a mean value with some weight function, generating Lorentz invariant theories and providing a connection with usual theories constructed in an ordinary $D=4$ spacetime.

To clarify a little bit what we saw in the last sections, we can say that, based on \cite{Amorim1,Amorim4,Amorim5,Amorim2} a new version of NCQM has been presented, where not only the coordinates ${\mathbf x}^\mu$ and their canonical momenta ${\mathbf p}_\mu$ are considered as operators in a Hilbert space
${\cal H}$, but also the objects of noncommutativity $\theta^{\mu\nu}$ and their canonical conjugate momenta $\pi_{\mu\nu}$.
All these operators belong to the same algebra and have the same hierarchical level, introducing  a minimal canonical
extension of the DFR algebra, i.e., the DFRA algebra introduced by R.~Amorim in a f\/irst paper~\cite{Amorim1} followed by others \cite{Amorim4,Amorim5,Amorim2,aa}.
This enlargement of the usual set of Hilbert space operators allows the theory to be invariant under the rotation group $SO(D)$, as showed
in detail in the sections above \cite{Amorim1,Amorim2}, when the treatment was a nonrelativistic one.  Rotation invariance in a nonrelativistic theory, is fundamental if one intends to
describe any physical system in a consistent way. In the last sections, the corresponding relativistic treatment was presented, which permits to implement Poincar\'{e} invariance as
a dynamical symmetry   \cite{Iorio} in  NCQM.%  \cite{NCQM}.
In this section we essentially consider the ``second quantization'' of the model discussed above~\cite{Amorim4}, showing that the extended Poincar\'e symmetry here is generated via generalized Heisenberg relations, giving the same algebra displayed in \cite{Amorim4,Amorim5}.

Now we will study the new NC charged Klein--Gordon theory described above in this $D=10$,  $x+\theta$ space and
analyze its symmetry structure, associated with the invariance of the action under some extended Poincar\'e (${\cal P}'$) group.  This symmetry structure is also displayed inside the second quantization level, constructed via generalized Heisenberg relations. After that, the f\/ields are shown as expansions in a plane wave basis in order to solve the equations of motion using the Green's functions formalism adapted for this new $(x+\theta)$ $D=4+6$ space.  It is assumed in this section that $\Omega(\theta)$ def\/ined in \eqref{bb8} is constant.

\subsection{The action and symmetry relations}

An important point is that, due to (\ref{a1}), the operator ${\mathbf x}^\mu$ can not
be used to label a possible basis in ${\cal H}$. However, as the components of
${\mathbf X}^\mu$ commute, as can be verif\/ied from the DRFA algebra and the relations following this one, their
eigenvalues  can be used for such purpose. From now on let us denote  by $x$ and
$\theta$ the eigenvalues of ${\mathbf X}$ and ${\mathbf\theta}$.

In Section~\ref{section4} we saw that the relations (\ref{a1}), \eqref{6a}--\eqref{12} and \eqref{15000} allowed us to utilize~\cite{Gracia}
\begin{gather*}
%\label{i16}
{ \mathbf M}^{\mu\nu}= { \mathbf X}^\mu{\mathbf p}^\nu-{\mathbf X}^\nu{\mathbf
p}^\mu-\theta^{\mu\sigma}\pi_\sigma^{\,\,\nu}+\theta^{\nu\sigma}\pi_\sigma^{\,\,\mu}
\end{gather*}
as the  generator of the Lorentz group, where   %\cite{NCQM}
\begin{gather*}
%\label{i5}
{\mathbf X}^\mu={\mathbf x}^\mu+{1\over2}{\mathbf\theta}^{\mu\nu}{\mathbf p}_\nu ,
\end{gather*}
and we see that the proper algebra is closed, i.e.,
\begin{gather*}
%\label{i17}
[{\mathbf M}^{\mu\nu},{\mathbf M}^{\rho\sigma}]=i\eta^{\mu\sigma}{\mathbf
M}^{\rho\nu}-i\eta^{\nu\sigma}{\mathbf M}^{\rho\mu}-i\eta^{\mu\rho}{\mathbf
M}^{\sigma\nu}+i\eta^{\nu\rho}{\mathbf M}^{\sigma\mu} .
\end{gather*}
Now $\mathbf M^{\mu\nu}$
generates the expected symmetry transformations when acting on all the operators in
Hilbert space.
Namely, by def\/ining the dynamical transformation of an arbitrary operator ${\mathbf A}$ in
$\cal H$ in such a way that
$\delta {\mathbf A}=i[ {\mathbf A}, {\mathbf G}]$, where
\begin{gather*}
{\mathbf G}={1\over2}\omega_{\mu\nu}{\mathbf M}^{\mu\nu}-a^\mu{\mathbf
p}_\mu+{1\over2}b^{\mu\nu}\pi_{\mu\nu} ,
\end{gather*}
and $\omega^{\mu\nu}=-\omega^{\nu\mu}$, $a^\mu$, $b^{\mu\nu}=-b^{\nu\mu}$ are
inf\/initesimal parameters, it follows that
\begin{subequations}
\label{i199}
\begin{gather}
\delta {\mathbf x}^\mu  =  \omega ^\mu_{\,\,\,\,\nu}{\mathbf x}^\nu+a^\mu+{1\over2}b^{\mu\nu}{\mathbf p}_\nu ,\label{i199a} \\
\delta {\mathbf X}^\mu  =  \omega ^\mu_{\,\,\,\,\nu}{\mathbf X}^\nu+a^\mu ,\label{i199b} \\
\delta{\mathbf p}_\mu  = \omega _\mu^{\,\,\,\,\nu}{\mathbf p}_\nu , \label{i199c} \\
\delta\theta^{\mu\nu} = \omega ^\mu_{\,\,\,\,\rho}\theta^{\rho\nu}+ \omega ^\nu_{\,\,\,\,\rho}\theta^{\mu\rho}+b^{\mu\nu} ,\label{i199d} \\
\delta\pi_{\mu\nu} = \omega _\mu^{\,\,\,\,\rho}\pi_{\rho\nu}+
\omega _\nu^{\,\,\,\,\rho}\pi_{\mu\rho} , \label{i199e} \\
\delta {\mathbf M}^{\mu\nu} = \omega ^\mu_{\,\,\,\,\rho}{\mathbf M}^{\rho\nu}+
\omega ^\nu_{\,\,\,\,\rho}{\mathbf M}^{\mu\rho}+a^\mu{\mathbf p}^\nu-a^\nu{\mathbf
p}^\mu+b^{\mu\rho}\pi_\rho^{\,\,\,\,\nu}+ b^{\nu\rho}\pi_{\,\,\,\rho}^{\mu} ,\label{i199f}
\end{gather}
\end{subequations}
 generalizing the action of the Poincar\'{e} group ${\cal P}$ in order to include
$\theta$ and $\pi$ transformations, i.e.,~${\cal P}'$. The ${\cal P}'$ transformations
close in an algebra, such that
\begin{gather*}%\label{xxx}
[\delta_2,\delta_1] {\mathbf A}=\delta_3 {\mathbf A} ,
\end{gather*}
and the parameters composition rule is given by{\samepage
\begin{gather*} \omega^\mu_{3\,\,\,\,\nu}=\omega^\mu_{1\,\,\,\,\alpha}\omega^\alpha_{2\,\,\,\,\nu}
-\omega^\mu_{2\,\,\,\,\alpha}\omega^\alpha_{1\,\,\,\,\nu} ,
\qquad
  a_3^\mu=\omega^\mu_{1\,\,\,\nu}a_2^\nu-\omega^\mu_{2\,\,\,\nu}a_1^\nu ,\nonumber\\
b_3^{\mu\nu}=\omega^\mu_{1\,\,\,\rho}b_2^{\rho\nu}-\omega^\mu_{2\,\,\,\rho}b_1^{\rho\nu}
-\omega^\nu_{1\,\,\,\rho}b_2^{\rho\mu}+
\omega^\nu_{2\,\,\,\rho}b_1^{\rho\mu} .%\label{i19bb}
\end{gather*}
The symmetry structure displayed in equation (\ref{i199}) was discussed before.}

In the last sections we tried to clarify these points with some detail and we have showed a~way for
constructing  actions representing possible f\/ield theories in this extended
$x+\theta$ spacetime.  One of such actions, generalized in order to permit the
scalar f\/ields to be complex, is given by
\begin{gather}
\label{b8}
S=-\int d^{4}\,x\,d^{6}\theta \left\{ \partial^\mu\phi^*\partial_\mu\phi +
 {{\lambda^2}\over4} \partial^{\mu\nu}\phi^*\partial_{\mu\nu}\phi
+m^2 \phi^*\phi\right\},
\end{gather}
where $\lambda$ is a parameter with dimension of length, as the Planck
length, which is introduced due to dimensional reasons. Here we are also suppressing
a possible  factor $\Omega(\theta)$ in the measure, which is a scalar weight
function, used in~\cite{Carlson,Haghighat,Carone,Ettefaghi,Morita,Saxell},
in a NC gauge
theory context,  to make the connection between the $D=4+6$
and the $D=4$ formalisms.

The corresponding Euler--Lagrange equation reads
\begin{gather}
\label{c6}
{{\delta S}\over{\delta\phi}}  =\big(\Box +\lambda^2\Box_\theta- m^2\big)\phi^*
  = 0 ,
\end{gather}
with a similar equation of motion for $\phi$. The action (\ref{b8}) is
invariant under the transformation
\begin{gather}
\label{AB19cc}
\delta \phi=-(a^\mu+\omega^\mu_{\,\,\,\nu}x^\nu)\,\partial_\mu\phi
-{1\over2}(b^{\mu\nu}+2\omega^\mu_{\,\,\,\rho}\theta^{\rho\nu})\,\partial_{\mu\nu}\phi ,
\end{gather}
besides the phase transformation
\begin{gather}
\label{i19dd}
\delta \phi=-i\alpha \phi ,
\end{gather}
with  similar expressions for $\phi^*$, obtained from (\ref{AB19cc}) and
(\ref{i19dd}) by complex conjugation.
 We observe that
  (\ref{i199c}) closes in an algebra, as in (\ref{i19aa}), with the same
composition rule def\/ined in~\eqref{i19b}. That equation def\/ines  how a complex
scalar f\/ield transforms in the $x+\theta$ space under~${\cal P}'$. The
transformation subalgebra generated by (\ref{i199d}) is of course Abelian, although
it could be directly generalized to a more general setting.

Associated with those symmetry transformations, we can def\/ine the conserved
currents~\cite{Amorim4}
\begin{gather*}
%\label{920}
 j^\mu={{\partial {\cal L}}\over{\partial\partial_\mu\phi}}\delta\phi+
\delta\phi^*{{\partial {\cal L}}\over{\partial\partial_\mu\phi^*}}+
{\cal L}\delta x^\mu ,\qquad
  j^{\mu\nu}={{\partial {\cal L}}\over{\partial\partial_{\mu\nu}\phi}}\delta\phi+
\delta\phi^*{{\partial {\cal L}}\over{\partial\partial_{\mu\nu}\phi^*}}
+{\cal L}\delta \theta^{\mu\nu} .
\end{gather*}
Actually, by using (\ref{i199c}) and (\ref{i199d}),  as well as (\ref{i199b})
and~(\ref{i199d}), we can show, after some algebra, that
\begin{gather*}%\label{94}
\partial_\mu j^\mu+\partial_{\mu\nu}j^{\mu\nu}=-{{\delta
S}\over{\delta\phi}}\delta\phi-\delta\phi^*{{\delta S}\over{\delta\phi^*}} .
\end{gather*}

The expressions above, as seen before, allow us to derive a specif\/ic charge
\begin{gather*}
%\label{950}
Q=-\int d^3 x d^6 \theta\, j^0 ,
\end{gather*}
for each kind of conserved symmetry encoded in (\ref{i199c}) and~(\ref{i199d}), since
\begin{gather*}
%\label{96}
\dot Q=\int d^3 x d^6 \theta\left(\partial_i j^i+{1\over2}\partial_{\mu\nu}j^{\mu\nu}\right)
\end{gather*}
vanishes as a consequence of the divergence theorem in this $(x,\theta)$ extended space. Let us consider each
specif\/ic symmetry  in~(\ref{i199c}) and~(\ref{i199d}). For usual $x$-translations, we can write
$j^0=j^0_\mu a^\mu$, and so we can def\/ine the total momentum
\begin{gather}
\label{95.1}
P_\mu  =  -\int d^3 x d^6 \theta\, j^0_\mu
  =  \int d^3 x d^6
\theta\big(\dot\phi^{*}\partial_\mu\phi+\dot\phi\partial_\mu\phi^{*}-{\cal
L}\delta^0_\mu\big) .
\end{gather}
 For $\theta$-translations, we can write that $j^0=j^0_{\mu\nu}b^{\mu\nu}$, and consequently,
giving
\begin{gather}
\label{96.1}
P_{\mu\nu}  =  -\int d^3 x d^6 \theta\, j^0_{\mu\nu}
  =  {1\over2}\int d^3 x d^6
\theta\big(\dot\phi^*\partial_{\mu\nu}\phi+\dot\phi\partial_{\mu\nu}\phi^*\big) .
\end{gather} In a similar way we def\/ine the Lorentz charge. By using the operator
\begin{gather}
\label{97}
\Delta_{\mu\nu}=x_{[\mu}\partial_{\nu]}+\theta_{[\mu}^{\,\,\,\alpha}\partial_{\nu]\alpha} ,
\end{gather}
 and def\/ining $j^0={\bar j}^0_{\mu\nu}\omega^{\mu\nu}$, we can write
\begin{gather}
\label{98}
M_{\mu\nu}   =  -\int d^3 x d^6 \theta\, {\bar j}^0_{\mu\nu}
  =  \int d^3 x d^6
\theta\big(\dot\phi^*\Delta_{\nu\mu}\phi+\dot\phi\Delta_{\nu\mu}\phi^*-{\cal
L}\delta^0_{[\mu}x_{\nu]}\big) .
\end{gather}

 At last, for the symmetry given by (\ref{i199d}), we can write the $U(1)$ charge as
\begin{gather}
\label{99}
{\cal Q}= i \int d^3 x d^6 \theta\,(\dot\phi^*\phi-\dot\phi\phi^*) .
\end{gather}

Now let us show that these charges generate the appropriate f\/ield transformations
(and dynamics) in a quantum scenario, as generalized Heisenberg relations.
To start the quantization of such theory, we can def\/ine as usual the f\/ield momenta
\begin{gather}
\label{100}
  \pi={{\partial{\cal L}}\over{\partial\dot\phi}}=\dot\phi^* ,\qquad
 \pi^*={{\partial{\cal L}}\over{\partial\dot\phi^*}}=\dot\phi ,
\end{gather}
satisfying the non vanishing equal time commutators (in what follows the
commutators are to be understood as equal time commutators)
\begin{gather}
[\pi(x,\theta),\phi(x',\theta')]=-i\delta^3(x-x')\delta^6(\theta-\theta'),\nonumber\\
 [\pi^*(x,\theta),\phi^*(x',\theta')]=-i\delta^3(x-x')\delta^6(\theta-\theta').\label{101}
\end{gather}

The strategy now is just to generalize the usual f\/ield theory and rewrite the
charges (\ref{95.1})--(\ref{99}) by eliminating the time derivatives of the f\/ields in
favor of the f\/ield momenta. After that we use~(\ref{101}) to dynamically generate
the symmetry operations. In this spirit, accordingly to~(\ref{95.1}) and~(\ref{100}),
the spatial translation is generated by
\begin{gather*}
%\label{102}
P_i=\int d^3 x d^6 \theta \big(\pi\partial_i\phi+\pi^*\partial_i\phi^*\big) ,
\end{gather*}
and it is trivial to verify, by using (\ref{101}), that
\begin{gather*}
%\label{103}
[P_i,{\cal Y}(x,\theta)]=-i\partial_i{\cal Y}(x,\theta) ,
\end{gather*}
where ${\cal Y}$ represents $\phi$, $\phi^*$, $\pi$ or $\pi^*$.
The dynamics is generated by{\samepage
\begin{gather*}
%\label{102.1}
P_0=\int d^3 x d^6
\theta\left(\pi^*\pi+\partial^i\phi^*\partial^i\phi+{{\lambda^2}\over4}\partial^{\mu\nu}\phi^*\partial_{\mu\nu}\phi
+m^2\phi^*\phi\right)
\end{gather*}
accordingly to (\ref{95.1}) and (\ref{100}).}

With the Lagrangian given in (\ref{b8}) we can have that
\begin{gather*}
\pi_{\mu\nu} = \frac{\partial {\cal L}}{\partial({\partial}^{\mu\nu}\phi)}  = \frac{\lambda^2}{4} {\partial}_{\mu\nu}\phi^{*},\qquad
\pi^{*}_{\mu\nu}\,=\,\frac{\partial {\cal L}}{\partial({\partial}^{\mu\nu}\phi^{*})}  = \frac{\lambda^2}{4} {\partial}_{\mu\nu}\phi
\end{gather*}
these are the canonical conjugate momenta in $\theta$-space, the $\theta$-momenta.  Together with $\pi$ and $\pi^*$ we have the complete momenta space.

At this stage it is convenient to
assume that classically
\[
\partial^{\mu\nu}\phi^*\partial_{\mu\nu}\phi\geq0
\]
to assure that the Hamiltonian $H=P_0$ is positive def\/inite.  This condition can also be written as
\[
\pi^{\mu\nu}\pi_{\mu\nu}^* \geq 0 ,
\]
 since we always have an even exponential for $\lambda$.
By using the fundamental
commutators (\ref{101}), the equations of motion (\ref{c6}) and the def\/initions
(\ref{100}), it is possible to prove the Heisenberg relation
\begin{gather*}
%\label{103.1}
[P_0,{\cal Y}(x,\theta)]=-i\partial_0{\cal Y}(x,\theta) .
\end{gather*}

The $\theta$-translations, accordingly to (\ref{96.1}) and (\ref{100}), are generated by
\begin{gather*}
%\label{104}
P_{\mu\nu}=\frac12\int d^3 x d^6
\theta \big(\pi\partial_{\mu\nu}\phi+\pi^*\partial_{\mu\nu}\phi^*\big) ,
\end{gather*}
and one obtains trivially  by (\ref{101}) that
\begin{gather*}
%\label{105}
[P_{\mu\nu},{\cal Y}(x,\theta)]=-i\partial_{\mu\nu}{\cal Y}(x,\theta) .
\end{gather*}

Lorentz transformations are generated by (\ref{98}) in a similar way. The spatial
rotations generator is given by
\begin{gather}
\label{106}
M_{ij}=\int d^3 x d^6 \theta \big(\pi\Delta_{ji}\phi+\pi^*\Delta_{ji}\phi^*\big) ,
\end{gather}
while the boosts are generated by
\begin{gather}
M_{0i} = {1\over2}\int d^3 x d^6 \theta\Big\{\pi^*\pi
x_i-x_0 \big(\pi\partial_i\phi+\pi^*\partial_i\phi^* \big) +  \pi \big(2\theta_{[i}^{\,\,\,\gamma}\partial_{0]\gamma}-x_0\partial_i \big)\phi  \nonumber\\
\phantom{M_{0i} =}{}  +  \pi^*\big(2\theta_{[i}^{\,\,\,\gamma}\partial_{0]\gamma}-x_0\partial_i \big)\phi^* + \big(\partial_j\phi^*\partial_j\phi+{{\lambda^2}\over{4}}\partial^{\mu\nu}\phi^*\partial_{\mu\nu}\phi
 + m^2\phi^*\phi\big)x_i\Big\} .\label{107}
\end{gather}
As can be verif\/ied in a direct way for (\ref{106}) and in a little more
indirect way for (\ref{107})
\begin{gather*}
%\label{108}
[M_{\mu\nu},{\cal Y}(x,\theta)]=i\Delta_{\mu\nu}{\cal Y}(x,\theta) ,
\end{gather*}
for any dynamical quantity ${\cal Y}$, where $\Delta_{\mu\nu}$ has been
def\/ined in~(\ref{97}). At last
we can rewrite~(\ref{99}) as
\begin{gather*}
%\label{109}
{\cal Q}= i\,\int d^3 x d^6 \theta \big(\pi\phi-\pi^*\phi^*\big) ,
\end{gather*}
 generating (\ref{i199d}) and its conjugate, and similar expressions for~$\pi$ and~$\pi^*$. So, the $\cal{P}$' and (global) gauge transformations can be generated by the action of the operator
\begin{gather*}
%\label{110}
{G}={1\over2}\omega_{\mu\nu}{M}^{\mu\nu}-a^\mu{P}_\mu+{1\over2}b^{\mu\nu}{P}_{\mu\nu}-\alpha {\cal Q}
\end{gather*}
 over the complex f\/ields and their momenta, by using the canonical commutation relations~(\ref{101}). In this way the $\cal{P}'$ and gauge transformations are generated as generalized Heisenberg relations. This is a new result that shows the consistence of the DFRA formalism. Furthermore, there are also four Casimir operators def\/ined with the operators given above, with the same form as those previously def\/ined at a f\/irst quantized perspective.
So, the structure displayed above is very similar to the usual one found
in ordinary quantum complex scalar f\/ields. We can go one step further,
by expanding the f\/ields and momenta in modes, giving as well some other prescription,  to def\/ine the
relevant Fock space, spectrum, Green's functions and all the basic structure related to free bosonic
f\/ields. In what follows we consider some of these issues and postpone others for forthcoming works.

\subsection{Plane waves and Green's functions}

In order to evaluate a little more the framework described in the last sections,
 let us rewrite the generalized charged Klein--Gordon action (\ref{b8}) with source terms as
\begin{gather}
\label{0.0}
S = -  \int d^4 x d^6 \theta \left\{\p^\mu\phi^* \p_\mu \phi  + {\lambda^2 \over 4} \p^{\mu\nu}\phi^* \p_{\mu\nu} \phi + m^2 \phi^* \phi  + J^* \phi + J \phi^* \right\} .
\end{gather}
 The corresponding equations of motion   are
\begin{gather}
\label{0.1}
\big(\Box +\lambda^2\Box_\theta- m^2\big)\phi(x,\theta) = J(x,\theta)
\end{gather}
 as well as its complex conjugate one.
 We have the following formal solution{\samepage
\begin{gather*}
%\label{0.2.1}
\phi(x,\theta) = \phi_{J=0}(x,\theta) + \phi_J (x,\theta),
\end{gather*}
 where, clearly, $\phi_{J=0}(x,\theta)$ is the source free solution
and $\phi_J (x,\theta)$ is the solution with $J\neq 0$.}

The Green's function for
(\ref{0.1}) satisf\/ies
\begin{gather}
\label{0.3}
\big(\Box +\lambda^2\Box_\theta- m^2\big) G(x-x',\theta-\theta')= \delta^4 (x-x') \delta^6 (\theta-\theta') ,
\end{gather}
where $\delta^4 (x-x')$ and $\delta^6 (\theta-\theta')$ are the Dirac's delta functions
\begin{gather}
\label{0.9}
\delta^4 (x-x') = {1\over (2\pi)^4} \int d^4 K_{(1)}  e^{iK_{(1)}\cdot (x-x')} ,
\\
\label {0.10}
\delta^6 (\theta-\theta') = {1\over (2\pi)^6} \int d^6 K_{(2)}  e^{iK_{(2)}\cdot (\theta-\theta')} .
\end{gather}

Now let us def\/ine in $D=10$
\begin{gather}
\label{0.7}
X = \left(x^\mu,{1\over \lambda} \theta^{\mu\nu}\right)
\end{gather}
 and
\begin{gather}
\label{0.8}
K = \big(K^\mu_{(1)},\lambda K^{\mu\nu}_{(2)}\big) ,
\end{gather}
where $\lambda$ is a parameter that carries the dimension of length, as said before.  From~(\ref{0.7}) and~(\ref{0.8}) we
write that
\[
K\cdot X = K_{(1)\mu} x^\mu + {1\over 2} K_{(2)\mu\nu} \theta^{\mu\nu} .
\]
The factor ${1\over 2}$
is introduced in order to eliminate repeated terms. In what follows it will also be considered that
\[
d^{10}K=d^4 K_{(1)}d^6 K_{(2)}
\]
and
\[
d^{10}X=d^4 x d^6 \theta .
\]

So, from (\ref{0.1}) and (\ref{0.3}) we formally have that
\begin{gather}
\label{0.4}
\phi_J (X) = \int d^{10} X'\,  G(X-X') J(X') .
\end{gather}

To derive an explicit form for the Green's function, let us expand $G(X-X')$ in terms of plane waves. Hence, we can write that
\begin{gather}
\label{0.12}
G(X-X') = {1\over (2\pi)^{10}} \int d^{10}K \,\tilde{G}(K) e^{iK\cdot (X-X')} .
\end{gather}
 Now, from (\ref{0.3}), (\ref{0.9}), (\ref{0.10}) and (\ref{0.12}) we obtain that,
\begin{gather*}
%\label{0.13}
\big( \Box +\lambda^2\Box_\theta- m^2 \big) \int {d^{10}K\over (2\pi)^{10}} \tilde{G} (K) e^{iK\cdot (x-x')}
 = \int {{d^{10} K}\over (2\pi)^{10}} e^{iK\cdot (x-x')}
\end{gather*}
giving the solution for $\tilde{G}(K)$ as
\begin{gather}
\label{0.14}
\tilde{G} (K) = - {1\over {K^2 + m^2}}
\end{gather}
where, from (\ref{0.8}),
\[
K^2 = K_{(1)\mu} K_{(1)}^\mu + {\lambda^2 \over 2} K_{(2)\mu\nu} K_{(2)}^{\mu\nu} .
\]

Substituting (\ref{0.14}) in (\ref{0.12}) we obtain
\begin{gather}\label{0.17}
G(x-x',\theta-\theta')  = {1\over (2\pi)^{10}}\int d^{9} K \int d K^0 {1\over {(K^0)^2 - \omega^2}} e^{iK\cdot (x-x')},
\end{gather}
where the ``frequency'' in the $(x+\theta)$ space is def\/ined to be
\begin{gather}
\label{0.18}
\omega = \omega(\vec{K}_{(1)},K_{(2)})
 = \sqrt{\vec{K}_{(1)}\cdot \vec{K}_{(1)} + {\lambda^2\over 2}K_{(2)\mu\nu} K_{(2)}^{\mu\nu}+m^2},
\end{gather}
which can be understood as the dispersion relation in this $D=4+6$ space.  We can see also,
from~(\ref{0.17}), that there are two poles $K^0 = \pm \omega$ in this framework. Of course
 we can construct an analogous solution for
$\phi^*_{J} (x,\theta)$.

In general, the poles of the Green's function can be interpreted as masses for the stable particles described by the theory.  We can see directly from equation~(\ref{0.18}) that the plane waves in the $(x+\theta)$ space
establish the interaction between the currents in this space and have energy given by $\omega(\vec{K}_{(1)},K_{(2)})$
since
\[
\omega^2=\vec{K}_{(1)}^2+{\lambda^2 \over 2}K_{(2)}^2 +m^2=K^2_{(1,2)}+m^2 ,
\]
where
\[
K^2_{(1,2)} = \vec{K}_{(1)}^2+{\lambda^2 \over 2}K_{(2)}^2 .
\]
So, one can say that the plane waves that mediate the interaction describe the propagation of  particles in a $x+\theta$ spacetime with a mass
equal to $m$.
We ask if we can use the Cauchy residue theorem in this new space to investigate the contributions of the poles in~(\ref{0.17}).
Accordingly to the point described above,  we can assume that the Hamiltonian is positive def\/inite and it is directly related to the hypothesis that
 $K^2_{(1,2)} =-m^2 <0$. However if  the observables are constrained to a four dimensional spacetime, due to some kind of compactif\/ication, the physical mass can have contributions from the NC sector. This point is left for a forthcoming work~\cite{NEXT}, when we will consider the Fock space structure of the theory and possible schemes for compactif\/ication.

For completeness, let us note that substituting (\ref{0.4}) and (\ref{0.17}) into~(\ref{0.0}), we arrive at the ef\/fective action
\begin{gather*}
%\label{0.32}
S_{\rm ef\/f} = - \int d^4 x  d^6 \theta  d^4 x'  d^6 \theta'\, J^* (X) \int \frac{d^9 K}{(2\pi)^{10}} \int dK^0
\frac{1}{(K^{0})^2 - \omega^2 + i\varepsilon} e^{iK\cdot(X-X')} J(X') ,
\end{gather*}
which could be obtained, in a functional formalism, after integrating out the f\/ields.

We can conclude this last section saying that the DFRA formulation reviewed here takes into account noncommutativity without destroying the symmetry content of the corresponding commutative theories. We expect that the new features associated with the objects of noncommutativity will be relevant at high energy scales. Even if excited states in the Hilbert space sector associated with noncommutativity can not assessed,  ground state  ef\/fects could  in principle be detectable.

To end this revision work we have considered in this section, complex scalar f\/ields using a~new framework where the object of noncommutativity
$\theta^{\mu\nu}$ represents independent degrees of freedom.We have started  from a f\/irst quantized formalism, where $\theta^{\mu\nu}$
and its canonical momentum~$\pi_{\mu\nu}$ are considered as operators  living in some Hilbert space.
This structure, which is compatible with the minimal canonical extension of the Doplicher--Fredenhagen--Roberts--Amorim (DFRA) algebra,
 is also invariant under an extended Poincar\'e group of symmetry, but keeping, among others, the usual Casimir invariant operators.
After that,
in a second quantized formalism perspective, we explained an explicit form for the extended Poincar\'e
generators and the same algebra of the f\/irst quantized description has been generated via generalized Heisenberg relations.

\subsection*{Acknowledgments}

ACRM and WO would like to thank CNPq (Conselho Nacional de Desenvolvimento Cient\'{\i}f\/ico e Tecnol\'ogico) for partial f\/inancial support, and AOZ would like to thank CAPES (Coordena\c{c}\~ao de Aperfei\c{c}oamento de Pessoal de N\' ivel Superior) for the f\/inancial support.  CNPq and CAPES are Brazilian research agencies.

%\pdfbookmark[1]{References}{ref}
\LastPageEnding

\end{document}